\documentclass[aps,pra,nofootinbib,showpacs, twocolumn]{revtex4}
\usepackage{mathrsfs}
\usepackage{amsfonts}
\usepackage{amssymb}
\usepackage{amsmath}
\usepackage{epsf,graphics,graphicx}
\usepackage{float}
\usepackage{subfigure}
\newcommand{\ket}[1]{|#1\rangle}
\newcommand{\bra}[1]{\langle #1|}

\begin{document}
\title{Robustness of non-adiabatic holonomic gates}
\author{Markus Johansson$^{1,2}$}
\email{cqtbemj@nus.edu.sg} 
\author{Erik Sj\"oqvist$^{1,2}$}
\email{erik.sjoqvist@kvac.uu.se}
\author{L. Mauritz Andersson$^3$}
\email{mauritza@kth.se} 
\author{Marie Ericsson$^2$}
\email{marie.ericsson@kvac.uu.se}
\author{Bj\"orn Hessmo$^1$}
\email{phyhbg@nus.edu.sg}
\author{Kuldip Singh$^1$}
\email{sciks@nus.edu.sg} 
\author{D. M. Tong$^4$}
\email{tdm@sdu.edu.cn}
\affiliation{$^1$Centre for Quantum Technologies, National University
of Singapore, 3 Science Drive 2, 117543 Singapore, Singapore \\
$^2$Department of Quantum Chemistry, Uppsala University, Box 518, 
Se-751 20 Uppsala, Sweden \\
$^3$Department of Applied Physics, KTH Royal Institute of Technology, SE-100 44 Stockholm, 
Sweden \\
$^4$Physics Department, Shandong University, Jinan, 250100, China}
\date{\today}
\begin{abstract}
The robustness to different sources of error of the scheme for non-adiabatic holonomic gates 
proposed in [New J. Phys. {\bf 14}, 103035 (2012)] is investigated. Open system effects as well as errors in the 
driving fields are considered. It is found that the gates can be made error resilient by using 
sufficiently short pulses. The principal limit of how short the pulses can be made is given by the 
breakdown of the quasi-monochromatic approximation. A comparison with the resilience of 
adiabatic gates is carried out.
\end{abstract}
\maketitle
\section{Introduction}
A central challenge for the realization of practically useful quantum computation is to find systems 
where the implementation of qubits and gate operations is resilient to perturbations due to instabilities 
in the setup and to open system effects. Several approaches to realize this resilience of quantum gates 
have been proposed, including the use of decoherence-free subspaces \cite{zanras}, noisless subsystems \cite{knille}, and topological quantum computation \cite{kita}. One of these approaches is holonomic quantum computation (HQC), using 
adiabatic evolution \cite{zanardi99}. This is a general procedure to build universal sets of gates 
that are robust to certain kinds of parametric noise \cite{jiannis3}, by using non-Abelian adiabatic geometric
phases.

One of the most studied implementations of adiabatic HQC is that of Duan {\it et al.} \cite{duan01}, 
which utilizes an array of trapped ions that can be manipulated by laser fields. Schemes to 
implement adiabatic holonomic gates have also been proposed in superconducting nanocircuits 
using Josephson junctions \cite{faoro03}, and in semiconductor quantum dots \cite{solinas03}. 
The gates in Refs. \cite{duan01,faoro03,solinas03} have turned out to be difficult to realize 
experimentally. One reason for this is the long run-time required for the desired parametric 
control associated with adiabatic evolution. The run-time must be long enough to minimize 
non-adiabatic corrections, but sufficiently short so that the error rate due to open system effects 
and instabilities of the setup is small. 

A scheme for HQC based on non-adiabatic non-Abelian geometric phases has been proposed in Ref. 
\cite{erik}. This scheme allows for universal quantum computation using gates that can be implemented 
rapidly compared to the adiabatic schemes, and therefore avoids the problems associated with a long
run-time. Non-adiabatic HQC has been combined with decoherence free subspaces
\cite{xu12} and applied to coupled quantum dots as well as molecular
magnets \cite{azimi12}. In this paper, we study some aspects of the robustness of one-qubit gates of the 
proposed non-adiabatic scheme to different kinds of errors and compare it to the robustness of 
the corresponding adiabatic gates based on the Duan {\it et al.} scheme. 

To realize quantum computation, it is necessary to bring the error rate per gate operation down 
below some threshold where error correction can be used to make the computation reliable. 
This threshold error rate has been estimated in different settings and by different authors to be 
somewhere between $10^{-6}$ and $10^{-2}$ \cite{gott,kit,pres,sene,knill}. There are many 
different sources of imperfections and their relative importance is specific to which system is used 
for the implementation of the gate. We therefore limit our study to some general sources of error 
that are typically encountered in a variety of implementations. Specifically, we study the sensitivity 
to decay and dephasing and to imperfect control of the external driving fields.  

Previously, it has been shown that adiabatic holonomic quantum gates are robust to first order 
against small random perturbations of the path in parameter space \cite{jiannis3}. Further analysis of robustness to parametric noise has been carried out in Refs. \cite{kuvshin,solinas1,buivi,cosmo}, and robustness to environmental interaction has been studied in Refs. \cite{ellinas,fuentes,trullo,florio06,parodi,florio3,parodi2,moll}. Moreover, the adiabatic holonomic gates are insensitive 
to the rate at which the evolution is driven as long as the adiabatic approximation is valid \cite{jiannis3}.

The outline of the paper is as follows. In Sec. \ref{ytrew}, we briefly review  non-adiabatic and 
adiabatic HQC proposed in Refs. \cite{erik} and \cite{duan01}, respectively. In Sec. 
\ref{pert}, we study the resilience of  one-qubit gates to decay of the excited state, dephasing, 
detuning, and incorrect parameters of the driving fields. The paper ends with the conclusions.

\begin{figure*}[ht]
\centering
\phantom{kkkkkkkkkkkkkkkkkkkkkkkkkkkkkk}\includegraphics[width=0.35\textwidth]{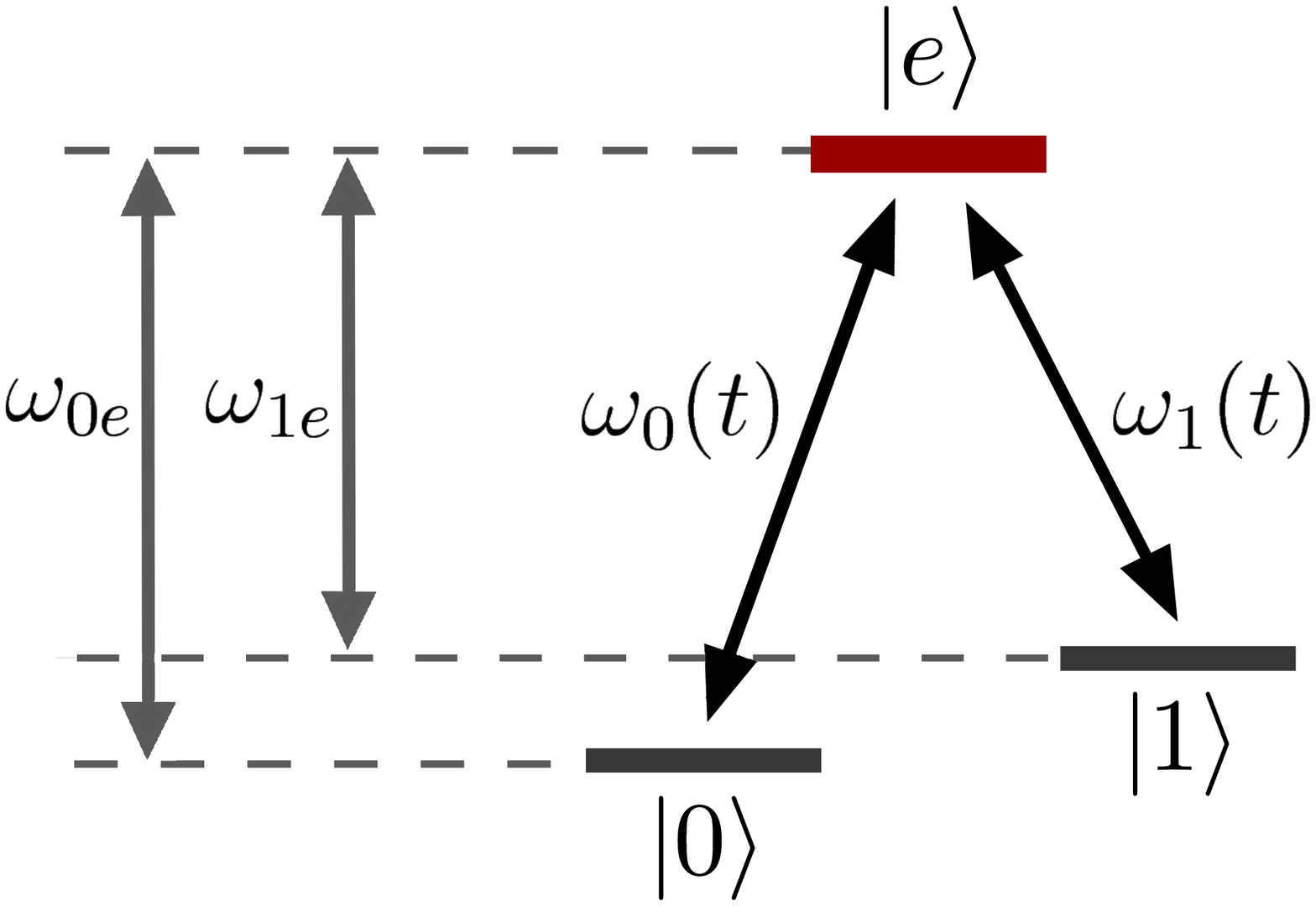}\phantom{kkkkkkkkkk}
\includegraphics[width=0.42\textwidth]{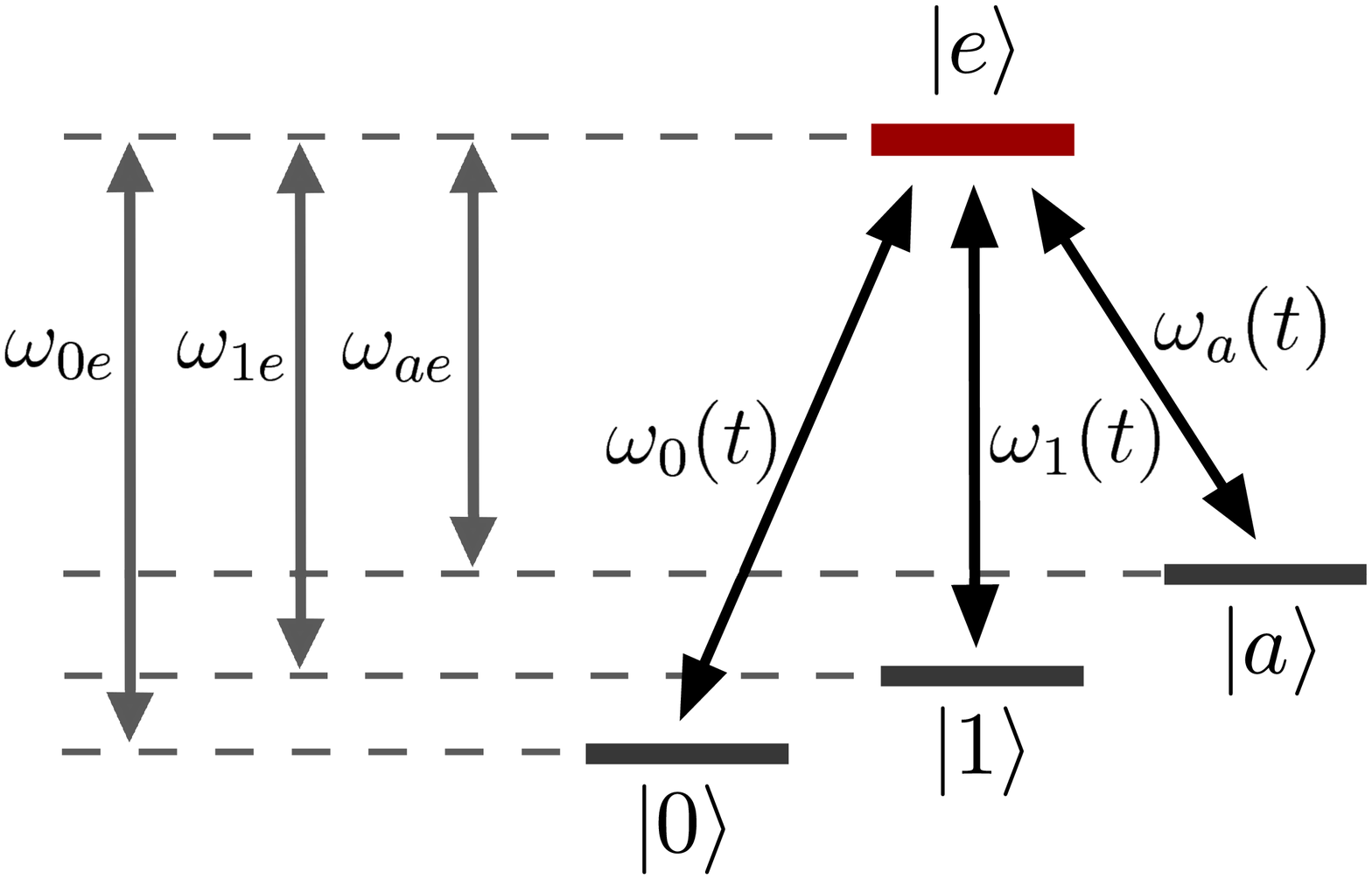}\phantom{.}
\caption{(Color online) Setup for the non-adiabatic single-qubit holonomic gate in a $\Lambda$ configuration (left) and for the adiabatic single-qubit holonomic gate in a tripod configuration (right). $\omega_{je}$ is the energy spacing of the $|j\rangle$ and $|e\rangle$ level, and the driving fields are $\omega_{j}(t)=\Omega(t)\omega_{j}e^{i\nu_{j}t}$ for $j=0,1$ in the non-adiabatic case and 
$\omega_{j}(t)=\Omega\omega_{j}(t/T)e^{i\nu_{j}t}$ for $j=0,1,a$ in the adiabatic case. The requirement for ideal gate implementation is $\nu_{j}=\omega_{je}$ in the non-adiabatic case and $\nu_0-\omega_{0e}=\nu_1-\omega_{1e}=\nu_a-\omega_{ae}$ in the adiabatic case. In the non-adiabatic setup the qubit is encoded in the $|0\rangle$ and $|1\rangle$ states. In the adiabtic setup the qubit is encoded in the dark subspace of the Hamiltonian. Initially the driving fields must therefore be chosen such that the dark subspace coincides with the  computational subspace spanned by $|0\rangle$ and $|1\rangle$. }
\label{fig:lambda}
\end{figure*}

\section{Holonomic gates}
\label{ytrew}

\subsection{The non-adiabatic gate}

The one-qubit non-adiabatic gate can be implemented in a $\Lambda$-system in the following 
way. A pair of zero-detuned pulses with the same real-valued pulse envelope $\Omega (t)$ couples 
selectively two ground state levels $\ket{0}$ and  $\ket{1}$ to an excited state $\ket{e}$. The 
corresponding Hamiltonian can be expressed in the rotating wave approximation as 
\begin{eqnarray}
H^{(\textrm{na})} (t) & = & \Omega(t) \left( \omega_0 \ket{e} \bra{0} +
\omega_1 \ket{e} \bra{1} + {\textrm{h.c.}} \right) .
\label{eq:ideal_na}
\end{eqnarray}
Here, the complex-valued driving parameters $\omega_0$ and $\omega_1$ satisfy 
$|\omega_0|^2 + |\omega_1|^2 = 1$, and describe the relative strength and relative phase 
of the $\ket{0} \leftrightarrow \ket{e}$ and $\ket{1} \leftrightarrow \ket{e}$ transitions. The 
Hamiltonian is turned on and off at $t=0$ and $t=\tau$, respectively, controlled by $\Omega (t)$. 
The pulse envelopes are described as monochromatic. It is therefore assumed that $\Omega(t)$ 
is slowly varying on the time scales $\frac{1}{\nu_j}$, where $\nu_j$ is the frequency of the 
driving field addressing the $\ket{j} \leftrightarrow \ket{e}$ transition, so that the quasi-monochromatic 
approximation is valid. This is equivalent to the requirement that $\frac{\Delta{\nu_j}}{\nu_j}\ll{1}$, 
where $\Delta{\nu_j}$ is the spectral width of the pulse \cite{mandel}. For the rotating wave 
approximation to be valid, it is also necessary that $\Omega(t)\ll\nu_j$ \cite{bloch}. Furthermore, 
we take $\ket{0}$ and $\ket{1}$ to define the one-qubit state space.
The $\Lambda$-configuration is illustrated in Fig. \ref{fig:lambda}.

The pulse-pairs are chosen such that $\omega_0$ and $\omega_1$ are time independent 
over the duration of each pulse pair. With this choice of parameters, the dark state 
$\ket{d} = -\omega_1 \ket{0} + \omega_0 \ket{1}$ decouples from the dynamics and 
the evolution is reduced to a simple Rabi oscillation between the bright state $\ket{b} = 
\omega_0^{\ast} \ket{0} + \omega_1^{\ast} \ket{1}$ and the excited state $\ket{e}$
\cite{fleischhauer96}. The Rabi frequency is $\Omega (t)$ and the subspace $M(t)$ 
spanned by $\ket{\psi_j (t)} = e^{-i\int_0^{t} H^{(\textrm{na})} (t') dt'} \ket{j} = U(t,0) \ket{j}$, 
$j=0,1$, undergoes a cyclic evolution if the pulse envelope satisfies $\int_{0}^{\tau} 
\Omega (t')dt' = \pi$. Under the above conditions, the final time evolution operator 
$U(\tau,0)$, projected onto the qubit space spanned by $\{ \ket{0}, \ket{1} \}$, defines 
the traceless Hermitian holonomic one-qubit gate
\begin{eqnarray}
U (C_{{\bf n}}) & = & {\bf n} \cdot \boldsymbol{\sigma} ,
\end{eqnarray}
where ${\bf n}=(\sin \theta \cos \phi , \sin \theta \sin \phi , \cos \theta )$ is a unit vector, 
$C_{{\bf n}}$ is the evolution corresponding to $\omega_0/\omega_{1}= -\tan (\theta /2) e^{i\phi}$,  
and $\boldsymbol{\sigma} = (\sigma_x,\sigma_y,\sigma_z)$ is a vector of the standard Pauli operators 
acting on $\ket{0},\ket{1}$. Thus, any traceless Hermitian SU(2) operation can be implemented 
using a pulse pair. Two pairs of pulses corresponding to the unit vectors ${\bf n}$ and ${\bf m}$ 
applied sequentially results in 
\begin{eqnarray}
U(C)&=&U(C_{{\bf m}}) U (C_{{\bf n}})\nonumber \\
 & = & {\bf m} \cdot {\bf n} + i \boldsymbol{\sigma} \cdot ({\bf m} \times {\bf n}) ,
\end{eqnarray}
which is an arbitrary SU(2) operation. Thus, $U(C)$ is a universal one-qubit gate.
The evolution is purely geometric since $\bra{\psi_j (t)} H^{(\textrm{na})}(t) \ket{\psi_k (t)} = 
\bra{j} H^{(\textrm{na})}(t) \ket{k} = 0$, $j,k=0,1$, for $t\in [0,\tau]$. $C$ can be interpreted 
as the path of $M(t)$ in the space of all 2-dimensional subspaces of the 3-dimensional Hilbert 
space, i.e., the Grassmann manifold $G(3;2)$.

\subsection{The adiabatic gate}
The adiabatic scheme proposed by Duan {\it et al.} \cite{duan01} is implemented utilizing a 
tripod-type system, where three ground states $\ket{0}$, $\ket{1}$, 
and $\ket{a}$ are coupled to an excited state $\ket{e}$ by the driving fields. Thus, the adiabatic 
implementation requires coherent control over an additional level compared to the non-adiabatic 
implementation using a $\Lambda$-system. The tripod configuration contains a 2-dimensional 
dark subspace, dependent on the field couplings. The system is prepared in a dark state belonging 
to the computational subspace spanned by $\ket{0}$ and $\ket{1}$, and the field couplings are varied 
independently such that the system remains approximately in an instantaneous dark state in the limit 
of large run-time $T$. More precisely, to remain in the adiabatic regime, the evolution in parameter 
space must be slow compared to the dynamical time scale of the system given by the energy 
difference between the dark and bright energy eigenstates. The Hamiltonian for this system in the rotating wave approximation is 
\begin{eqnarray}
\label{uyt}
H^{(\textrm{a})} (t) & = & 
\Omega \big[ \omega_0 (t/T) \ket{e} \bra{0}+\omega_1 (t/T) \ket{e} \bra{1} 
\nonumber\\
 & & + \omega_a (t/T)\ket{e}\bra{a}+{\textrm{h.c.}}\big]  , 
\label{eq:ideal_a}
\end{eqnarray}
where we have assumed zero-detuned field couplings. For ideal gate implementation it is however only necessary that the field couplings are equally detuned. The tripod configuration is illustrated in Fig. \ref{fig:lambda}.
The field parameters are varied through a closed loop in parameter space. 
In this way, a universal set of one-qubit gates consisting of $U_{1}(\Gamma_{1}) = 
e^{i\frac{1}{2} \Gamma_{1} |1\rangle\langle1|}$ and $U_{2}(\Gamma_{2}) = 
e^{i\Gamma_{2}\sigma_{y}}$, $\Gamma_{1}$ and $\Gamma_{2}$ being real numbers, 
can be realized. 

The $U_{1}$ gate can be implemented by choosing $\omega_0=0$, $\omega_1 = 
-\sin ( \vartheta / 2)e^{i\varphi}$, and $\omega_a=\cos ( \vartheta /2)$ and taking 
$\vartheta$ and $\varphi$ around a loop $c_1$ in parameter space enclosing the solid angle 
$\Gamma_{1} = \int_{c_1} \sin \vartheta d\vartheta d\varphi$. The $U_{2}$ gate can be 
implemented by choosing 
$\omega_0 = \sin \vartheta \cos \varphi $, $\omega_1= \sin \vartheta \sin \varphi$, and 
$\omega_a = \cos \vartheta$ and taking $\vartheta$ and $\varphi$ around a loop $c_2$ in 
parameter space enclosing the solid angle $\Gamma_{2}=\int_{c_2} \sin \vartheta d\vartheta 
d\varphi$.

\section{Analysis of gate robustness}

\label{pert}

There are two qualitatively different sources of error for the holonomic gates. One is open system 
effects caused by the interaction with the environment. The other is imperfect control of the 
parameters of the Hamiltonian due to unavoidable instabilities of the setup. These parameters 
are associated with the external driving fields used to control the system and are described as 
classical fields. 

Among open system effects, decay of the excited state and dephasing are important. These will 
be considered in Sec. (\ref{la}). The non-adiabatic scheme populates the excited state during 
the action of each pulse pair and it is therefore crucial to study the resilience to decay as a 
function of the operation parameters. In the adiabatic implementation, the excited state is 
populated only infinitesimally and therefore the adiabatic gates are robust against decay in 
the adiabatic limit. While decay is typically the dominant open system effect 
in trapped atomic systems \cite{brennen}, dephasing plays a central role in superconducting 
circuits \cite{naka,koch}.

Important sources of errors due to parametric instability are detuning and imperfect control of 
the parameters of the driving field. These will be considered in Sec. (\ref{lb}). Errors in detuning 
and pulse area in the non-adiabatic Abelian case for a two-level system have been analyzed in 
Ref. \cite{thom}.

The gate performance under open-system effects and parameter errors is quantified in terms 
of the gate fidelity $\bra{\chi} U^{\dagger}(C) \varrho_{{\textrm{out}}} U(C) \ket{\chi}$, where 
$U(C)$ is the desired gate operation and $\varrho_{{\textrm{out}}}$ is the output state computed 
from the dynamics for the non-adiabatic and adiabatic gates with error sources. For 
the open system effects and detuning, the gate fidelities are computed numerically for $4000$
input states $\ket{\chi}$, uniformly distributed over the Bloch sphere with respect to the Haar measure. The evolution of each input state is numerically integrated from the dynamical equations using the adaptive time step fourth order Runge-Kutta-Fehlberg (RKF45) method. In the numerical simulations 
the robustness to error sources is investigated using two test gates, the $\frac{\pi}{2}$ phase-shift 
gate $\ket{j} \mapsto e^{ij\pi/2} \ket{j}$, $j=0,1$, and the Hadamard gate $\ket{j} \mapsto
\frac{1}{\sqrt{2}} [(-1)^j \ket{j} + \ket{j \oplus 1}]$, $j=0,1$. Non-adiabatic and adiabatic 
implementations of these gates are described in Sec. (\ref{testg}).

\subsection{Test gates}
\label{testg}

\subsubsection{Non-adiabatic test gates}
In the non-adiabatic scheme, the $\frac{\pi}{2}$ phase-shift gate can be 
implemented by two pulse pairs, with the choice ${\bf n} = (\cos \phi,\sin \phi,0)$ and 
${\bf m} = (\cos \phi',\sin \phi',0)$. The Hadamard gate, can be implemented by a single 
pulse pair with ${\bf n} = \frac{1}{\sqrt{2}}(1,0,1)$. 

The $\pi$-pulses used in the numerical simulation are hyperbolic secant pulses with maximal 
amplitude $\beta$. Explicitly, for the $\frac{\pi}{2}$ phase-shift gate we choose the two pulse 
pairs as $\Omega (t) \left( \omega_0 ,\omega_1 \right) = \beta {\textrm{{sech}}} 
(\beta t) (-1,1)/\sqrt{2}$ and $\Omega (t-t_{s})\left( \omega_0' ,\omega_1' \right) = 
\beta {\textrm{{sech}}} [ \beta (t-t_{s})] (-1,e^{-i\pi/4}) / \sqrt{2}$, where $t_{s}$ is their temporal 
separation. For a given $\beta$, the temporal separation $t_{s}$ must be 
chosen large enough so that the pulse overlap is negligible. This is a necessary condition to avoid 
any spurious dynamical contributions to the gate. The Hadamard gate is realized by a single 
pulse pair with shape $\Omega (t) \left( \omega_0 ,\omega_1 \right) = \beta {\textrm{{sech}}} 
(\beta t) \left( 1,(\sqrt{2}-1)\right)/\sqrt{2(2-\sqrt{2})}$.

\begin{figure}[t]
\centering
\includegraphics[width=0.45\textwidth]{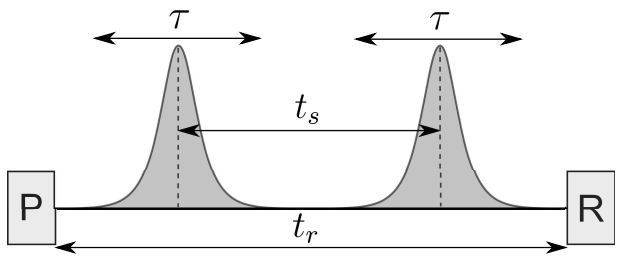}\phantom{.}
\caption{(Color online) Schematic illustration of the implementation of a non-adiabatic holonomic gate. The time $t_{r}$ 
between preparation (P) and read-out (R) of the qubit, the separation $t_{s}$ of the pulse-pairs 
and the pulse length $\tau$ are indicated.}
\label{fig:drawing}
\end{figure}

The pulses are truncated where the amplitude is $\beta/1000$, which gives the pulses a 
length $\tau=2\frac{1}{\beta}\textrm{arcsech} (\frac{1}{1000})$. Therefore, the pulse duration 
in the non-adiabatic setting decreases with increasing $\beta$ as a result of the pulse area 
being set to the fixed value $\pi$. This means that the total time $t_{r}$ between preparation 
and read-out can be decreased as well. Thus, by increasing $\beta$ we effectively speed up 
the gate. The implementation of two pulse pairs and the relevant operation times $\tau$, $t_{s}$, 
and $t_{r}$ are illustrated in Fig. \ref{fig:drawing}. Note that $\tau < t_s$ in order to avoid overlap of the pulses. 

\subsubsection{Adiabatic test gates}

Next, we consider the adiabatic scheme. The $\frac{\pi}{2}$ phase-shift gate can be implemented 
by one loop in parameter space, while the Hadamard gate requires two loops. The ideal adiabatic 
$\frac{\pi}{2}$ phase-shift gate is generated 
in the $\Omega{T} \rightarrow \infty$ limit by varying the field couplings  $\omega_0 = 0$, 
$\omega_1 = -\sin (\vartheta /2) e^{i\varphi}$ and $\omega_a = \cos (\vartheta /2)$ along 
the loop $(\vartheta,\varphi) = (0,0) \rightarrow (\frac{\pi}{2},0) \rightarrow (\frac{\pi}{2},\pi) 
\rightarrow (0,\pi) \rightarrow (0,0)$ at constant speed. Here, $|0\rangle$ is decoupled from 
the excited state.

\begin{figure}[t]
\centering
\includegraphics[width=0.45\textwidth]{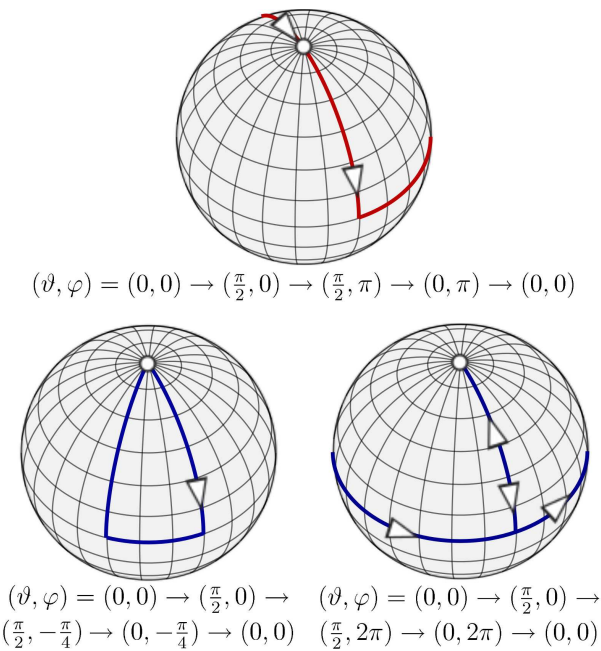}\phantom{.}
\caption{(Color online) The curve in parameter space used to implement the adiabatic holonomic $\frac{\pi}{2}$ 
phase-shift gate (top) and the two curves in parameter space used to implement the Hadamard 
gate (bottom). The top and right bottom curves correspond to the field couplings $\omega_0 = 0$, 
$\omega_1 = -\sin (\vartheta /2) e^{i\varphi}$ and $\omega_a = \cos (\vartheta /2)$, while the 
left bottom curve corresponds to $\omega_0 = \sin \vartheta \cos \varphi$, 
$\omega_1 = \sin \vartheta  \sin \varphi$ and $\omega_a = \cos \vartheta$.}
\label{spheres}
\end{figure}

\begin{figure*}[ht]
\centering
\includegraphics[width=0.44\textwidth]{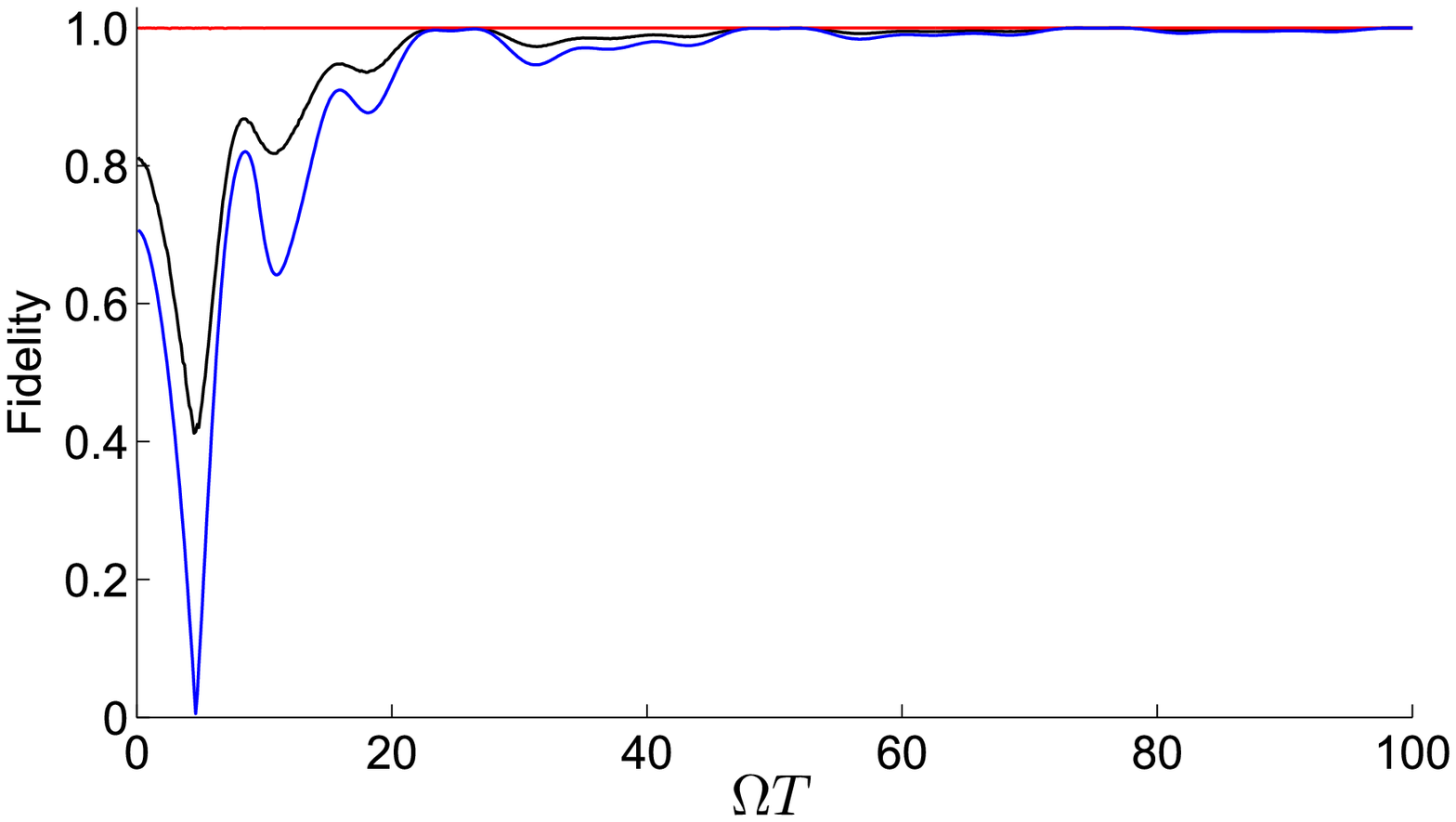}
\includegraphics[width=0.4\textwidth]{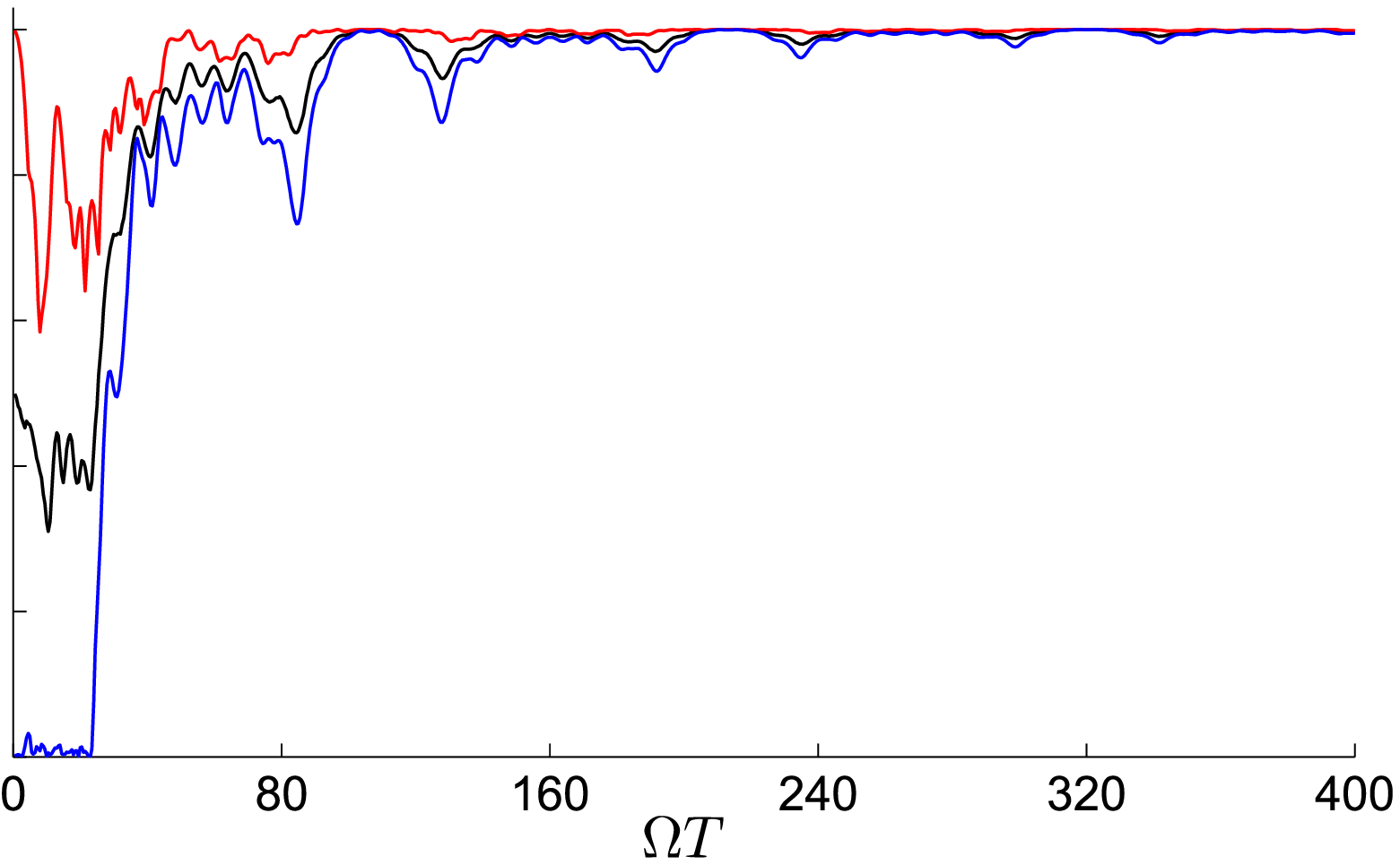}
\caption{(Color online) Influence of non-adiabatic corrections on the adiabatic holonomic 
$\frac{\pi}{2}$ phase-shift  gate (left) and Hadamard gate (right). The effect is quantified from top to bottom in 
terms of maximum (red), average (black), and minimum (blue) fidelities. We plot the fidelities 
as functions of the dimensionless quantity $\Omega T$, where $\Omega$ is the time independent 
global strength of field couplings and $T$ is the run-time of the gate.}
\label{graphs0}
\end{figure*}

The ideal Hadamard gate  is generated in the $\Omega{T} \rightarrow \infty$ limit by combining 
two loops. In the first loop, the field couplings are $\omega_0 = \sin \vartheta \cos \varphi$, 
$\omega_1 = \sin \vartheta  \sin \varphi$ and $\omega_a =  \cos \vartheta$ and these 
are varied along the loop $(\vartheta,\varphi) = (0,0) \rightarrow (\frac{\pi}{2},0) \rightarrow 
(\frac{\pi}{2},-\frac{\pi}{4}) \rightarrow (0,-\frac{\pi}{4}) \rightarrow (0,0)$ at constant speed.
In the second loop, the field couplings are  $\omega_0 = 0$, $\omega_1 =-\sin (\vartheta /2) 
e^{i\varphi}$ and $\omega_a = \cos (\vartheta /2)$ and these are varied along the loop
$(\vartheta,\varphi) = (0,0) \rightarrow (\frac{\pi}{2},0) \rightarrow (\frac{\pi}{2},2\pi) 
\rightarrow (0,2\pi) \rightarrow (0,0)$. The curves in parameter space used to implement the 
adiabatic gates are illustrated in Fig. \ref{spheres}.

Since the run-time $T$ and the strength of the field coupling $\Omega$ are always finite 
for adiabatic gates, there will be non-adiabatic corrections that reduce the fidelity. These 
corrections are due to failure of the state to remain in the dark subspace during evolution. In Fig. 
\ref{graphs0}, we show the fidelity for the adiabatic test gates, for finite $\Omega$ and ${T}$. 
The fidelities are plotted as functions of the dimensionless quantity $\Omega T$.  

The oscillatory behavior of the fidelities, as a function of $\Omega{T}$, is due to non-adiabatic effects. These effects can be understood as a combination of modified holonomies, caused by changes in the path of the computational subspace, and non-zero dynamical phases.
The revivals seen in the fidelities have been pointed out previously in Ref. \cite{florio06}. Some of 
the revivals reach unit fidelity and these can therefore be used to implement the desired gates for finite $\Omega{T}$. However in the non-adiabatic regime the gate is not holonomic since there are both dynamical and geometrical contributions to the gate operation. 
For the $\frac{\pi}{2}$ phase-shift gate the maximal fidelity (red line) is unity for all $\Omega{T}$ since $|0\rangle$ 
is decoupled from the dynamics.

\subsection{Open system effects}
\label{la}
\subsubsection{Decay}

To investigate the sensitivity of the non-adiabatic scheme to decay we assume that the time scale 
of the applied pulses is large compared to the time scale of the dynamics underlying the decay 
process, so that the Markovian approximation is valid. Given this, we further assume that the 
excited state decays to an auxiliary ground state level $\ket{g}$ \cite{foot} with a time independent rate 
$\gamma$. We compare the sensitivity of the non-adiabatic implementations of the test gates 
with that of the corresponding adiabatic implementations 
of the gates. The decay is modelled by the Lindblad equation
\begin{eqnarray}
\dot{\varrho}_t = -i[H(t),\varrho_t] + 2L \varrho_t L^{\dagger} - L^{\dagger} L \varrho_t -
\varrho_t L^{\dagger} L ,
\label{eq:lindblad}
\end{eqnarray}
where $\varrho_t$ is the density operator, $L = \sqrt{\gamma} \ket{g} \bra{e}$, and $H(t)$ is 
either $H^{(\textrm{na})} (t)$ or $H^{(\textrm{a})} (t)$. 

In the non-adiabatic case, the constraint of having $\pi$-pulses implies that the only experimentally 
controllable parameters of principal importance are the maximal coupling strength $\beta$ and the 
total time between preparation and read-out $t_{r}$. The dynamics of the gate can be described in 
terms of two dimensionless parameters. These can be chosen as $\beta / \gamma$ and 
$\gamma{t_{r}}$. When the pulse is a perfect $\pi$-pulse the excited state is negligibly populated 
after the end of the pulse. Therefore, the relevant time scale for decay is the width of the pulse 
$\tau$, which is proportional to $\frac{1}{\beta}$. Since $\gamma{\tau}\propto\frac{\gamma}{\beta}$ 
there is only one dynamically relevant dimensionless parameter.

The relevant operation parameters in the adiabatic case are the coupling strength $\Omega$ and 
run-time $T$. The dynamics can be characterized by two dimensionless parameters, which 
can be chosen as $\Omega{T}$ and $\gamma{T}$. In the simulations we consider both a fixed 
coupling strength $\Omega_{0}$ and vary $T$, and a fixed run-time $T_{0}$ and vary $\Omega$.

\begin{figure*}[htb!]
\centering
\includegraphics[width=0.325\textwidth]{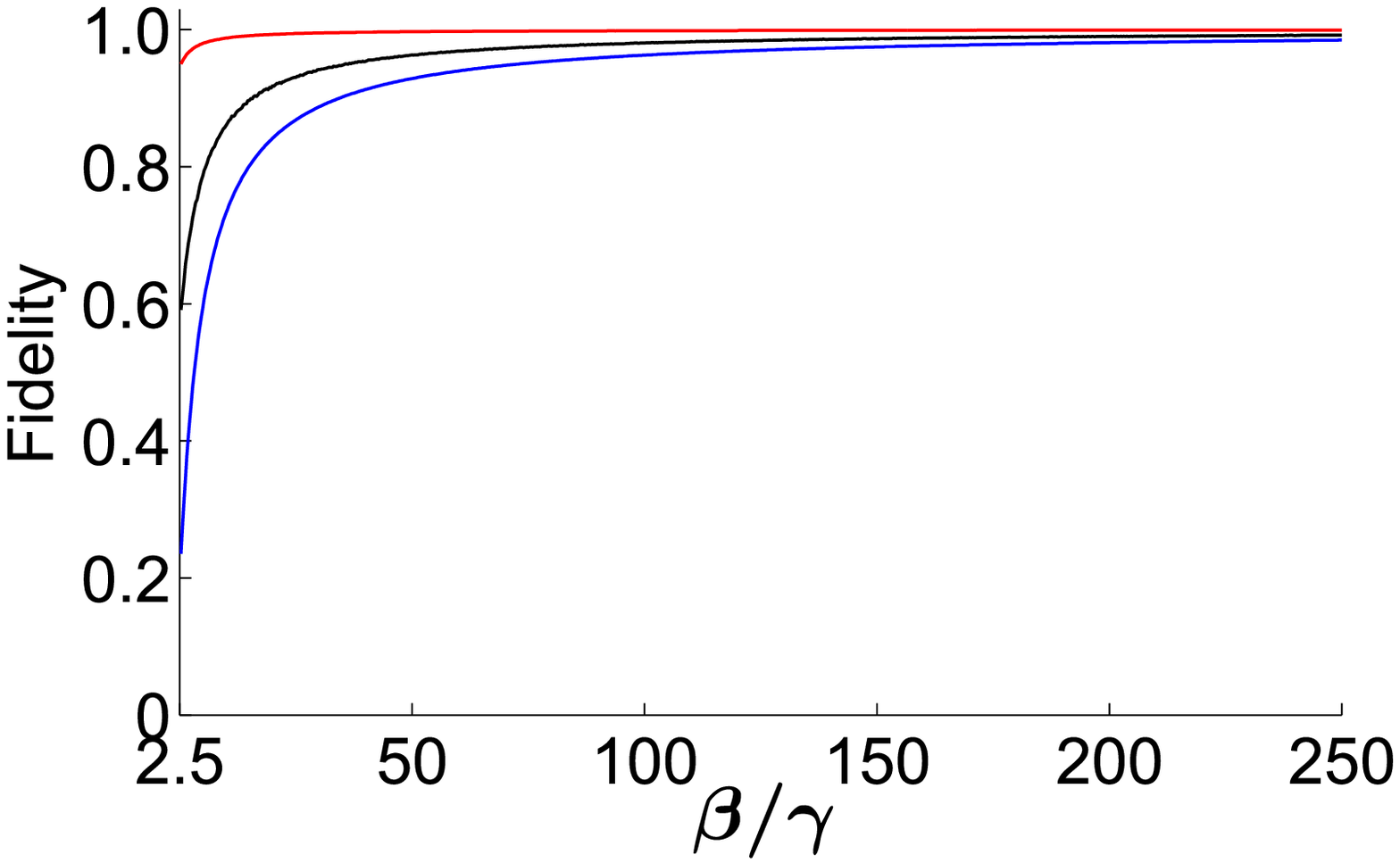}\phantom{.}
\includegraphics[width=0.32\textwidth]{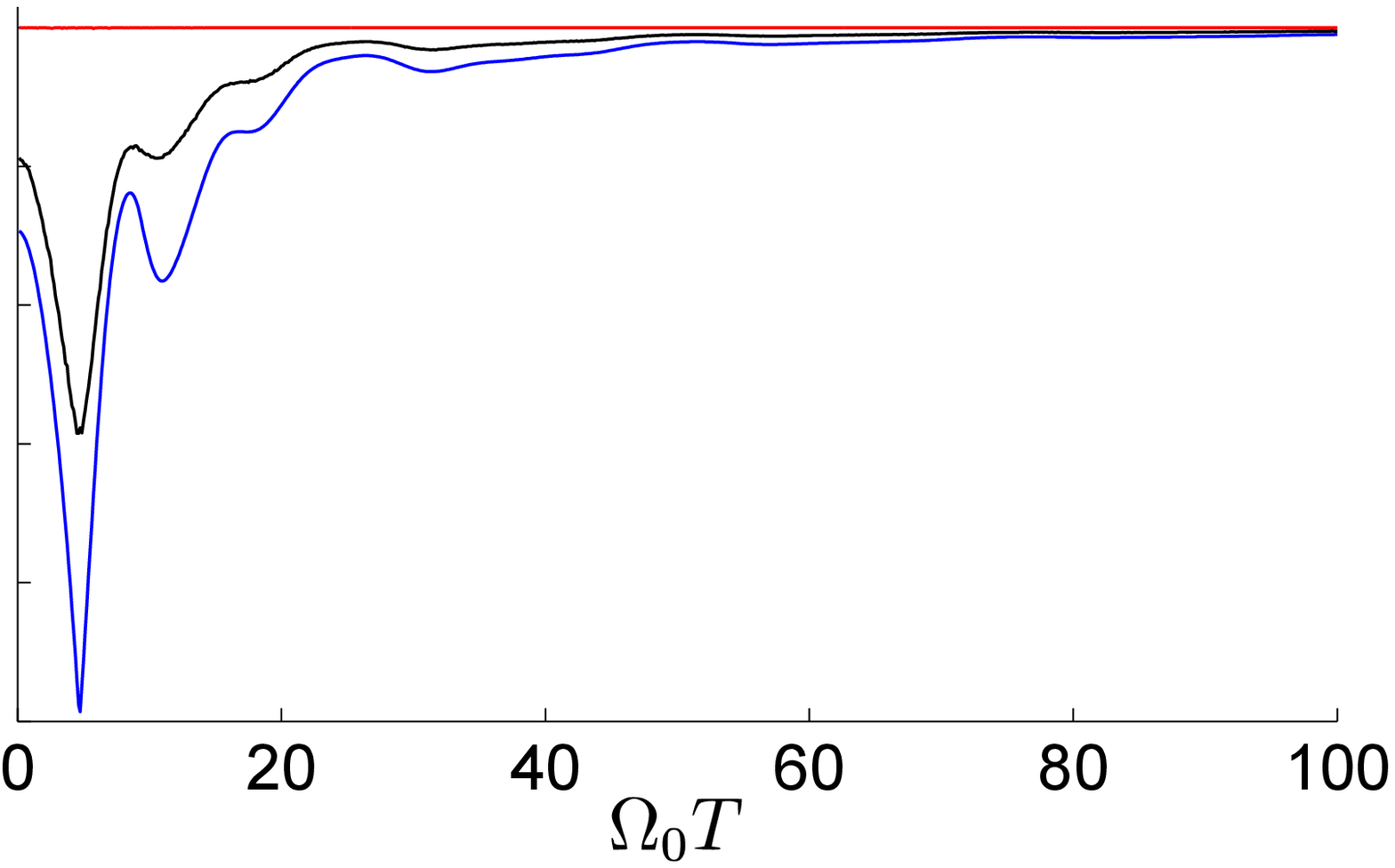}\phantom{.}
\includegraphics[width=0.32\textwidth]{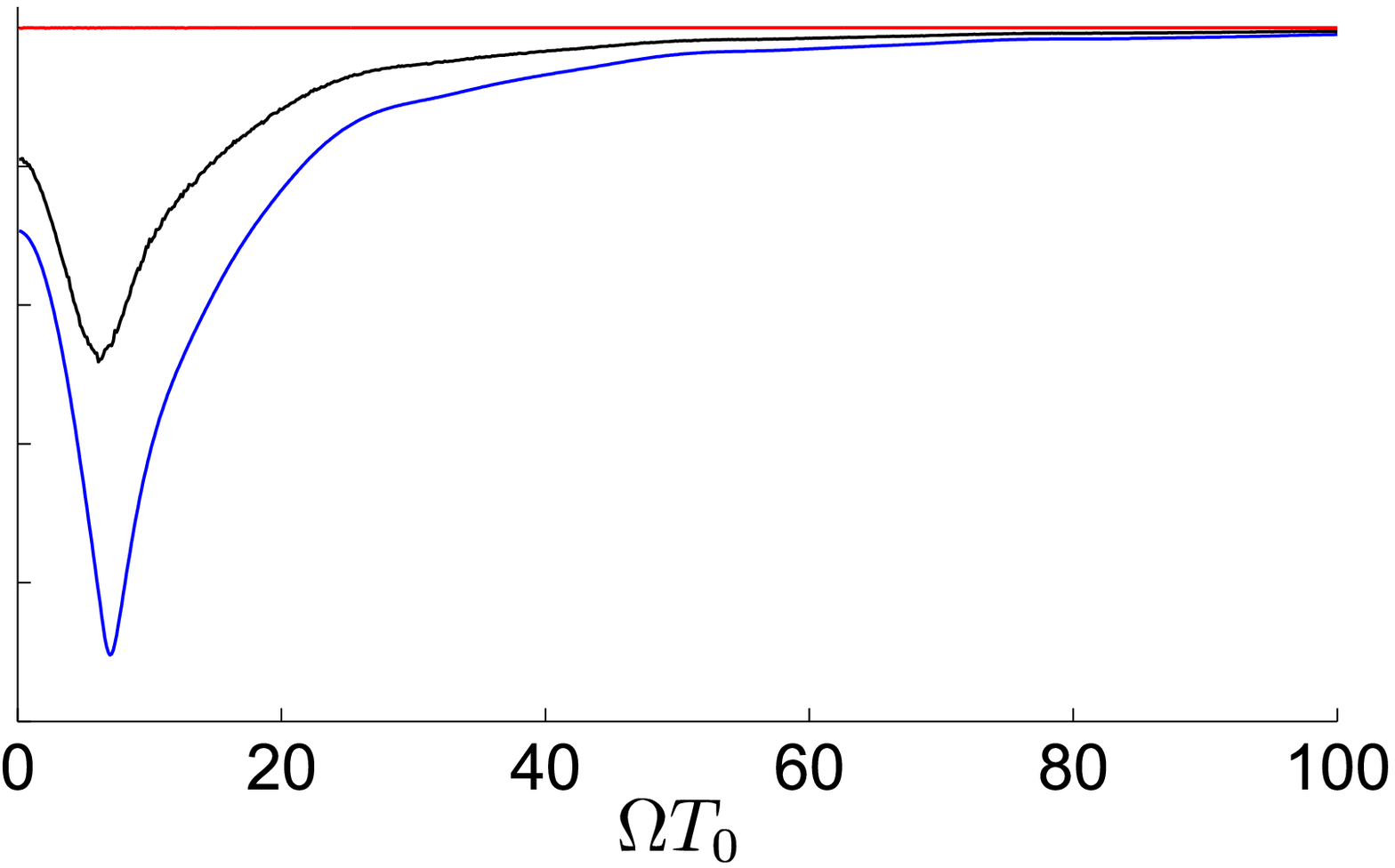}\phantom{.}
\centering
\includegraphics[width=0.325\textwidth]{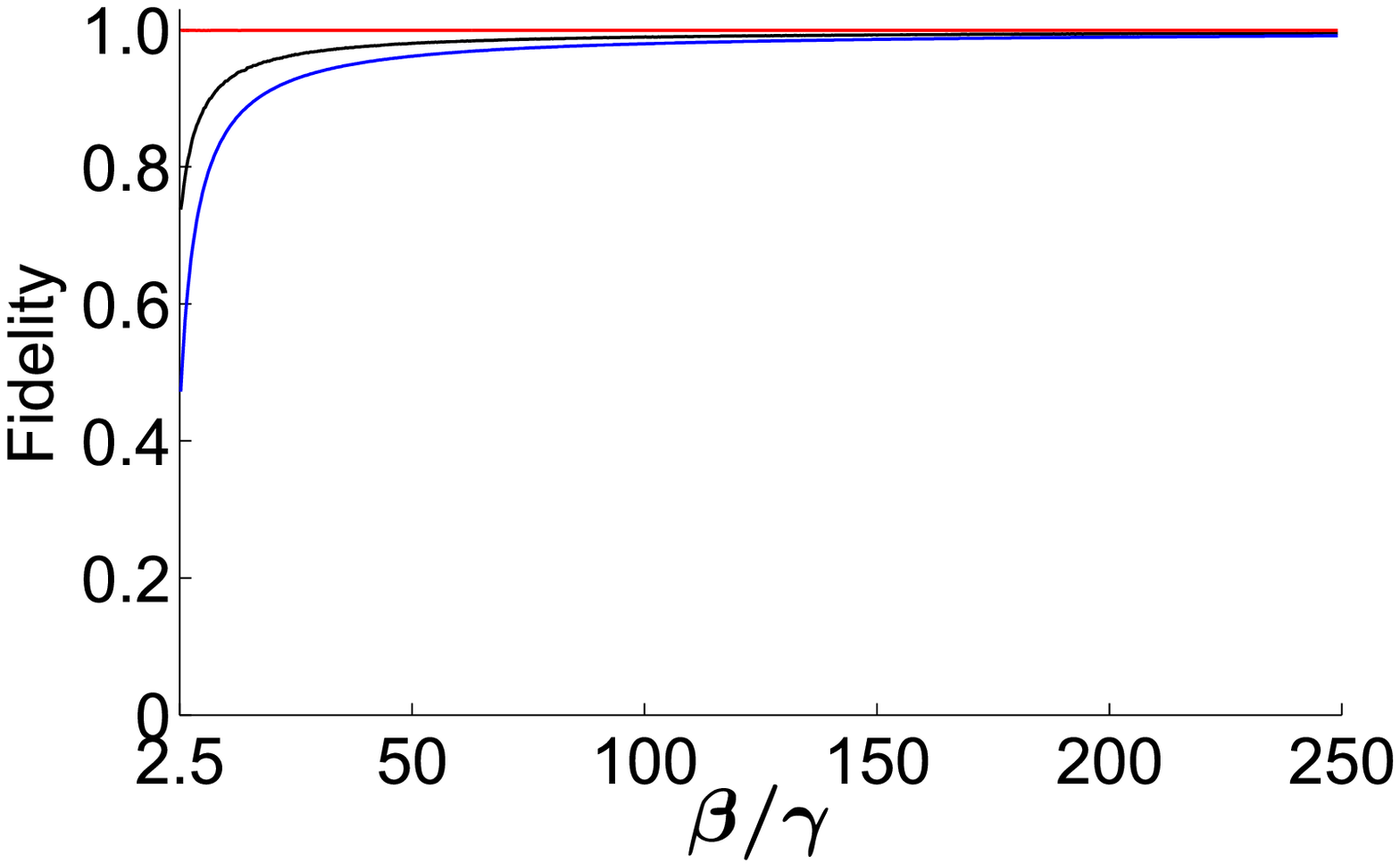}\phantom{.}
\includegraphics[width=0.32\textwidth]{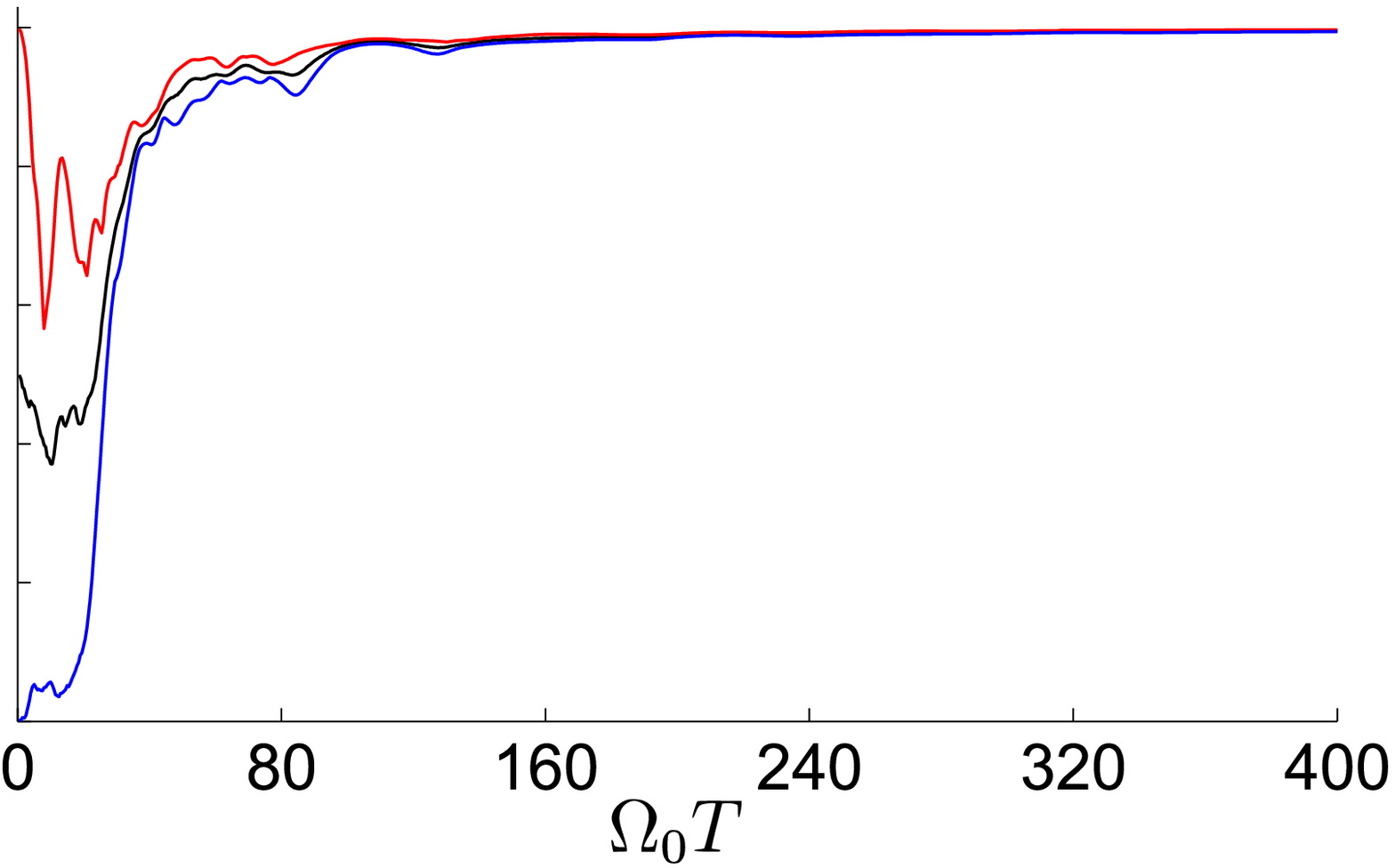}\phantom{.}
\includegraphics[width=0.32\textwidth]{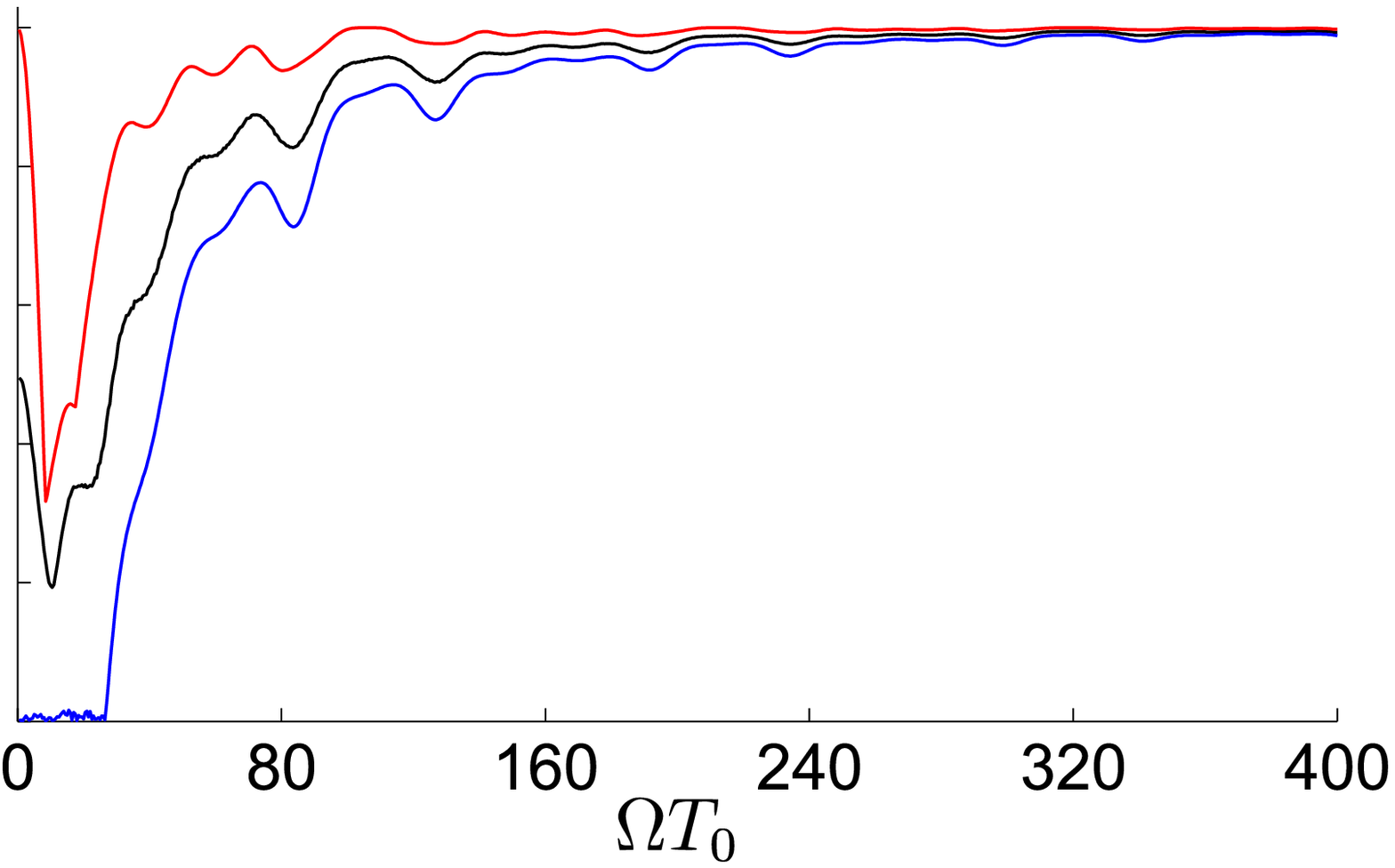}\phantom{.}
\caption{(Color online) Influence of decay with rate $\gamma$ of the excited state $|e\rangle$ 
on the non-adiabatic and adiabatic holonomic $\frac{\pi}{2}$ phase-shift gate (upper) and 
Hadamard gate (lower). The effect is quantified from top to bottom in 
terms of maximum (red), average (black), and minimum (blue) fidelities. The three panels show from left to right, the non-adiabatic gate 
with decay (left) and the adiabatic gate with decay, for a fixed coupling strength (middle) and 
fixed run-time (right). Choosing hyperbolic secant $\pi$-pulses with amplitude $\beta$, 
the non-adiabatic fidelities are plotted as functions of the dimensionless quantity $\beta /\gamma$. 
We plot the adiabatic fidelities as functions of the dimensionless quantities $\Omega_{0} T$ and $\Omega T_{0}$, where 
$\Omega$ is the time independent global strength of field couplings, $T$ is the run-time 
of the gate, and  $\Omega_{0}$ and $T_{0}$ are particular fixed values of these quantities. For the adiabatic gates we have chosen $\Omega_{0} /\gamma = 12.5$ for the case with 
fixed coupling strength and for the case of fixed run-time we have chosen $\gamma{T_{0}}=8$ 
for the $\frac{\pi}{2}$ phase-shift gate and $\gamma{T_{0}}=32$ for the Hadamard gate. In the 
non-adiabatic case, we have chosen $\gamma t_{s} = 8$, where $t_{s}$ is the temporal separation 
of the two pulses. $t_{s}$ is chosen sufficiently large to avoid pulse overlap for the $\beta /\gamma$
range shown in the left panels.}
\label{fig:graphs}
\end{figure*}

In Fig. \ref{fig:graphs}, we show the fidelities of the test gates, computed from Eq. (\ref{eq:lindblad}). 
The fidelities are plotted as functions of the dimensionless quantity $\beta/\gamma$ in the non-adiabatic case, as well as $\Omega_{0} T$ and $\Omega T_{0}$ in the adiabatic case for fix coupling strength and fixed run-time, respectively. For the adiabatic case, we 
have chosen $\Omega_{0} /\gamma = 12.5$, in the case of fixed coupling strength, and 
in the case of fixed run-time we have chosen $\gamma{T_{0}}=8$ for the $\frac{\pi}{2}$ 
phase-shift gate and $\gamma{T_{0}}=32$ for the Hadamard gate. These choices are made to 
make the effect of decay non-negligible. Furthermore, in the non-adiabatic case we have 
chosen $\gamma t_{s} = 8$, which guarantees that there is no pulse overlap for the 
$\beta /\gamma$ range shown.

The fidelities of the non-adiabatic gates tend monotonically to unity in the large $\beta /\gamma$
limit for both test gates (left panels). This demonstrates that the non-adiabatic holonomic 
test gates can be made robust to decay of the excited state by employing pulses that are sufficiently short compared to the decay time $\frac{1}{\gamma}$. 
For the Hadamard gate the maximal fidelity (red line) is identity for all $\beta/\gamma$. This is 
because it is implemented by a single pulse and the dark state of the Hamiltonian corresponding 
to this pulse is left unchanged by the dynamics and hence unaffected by the decay. For the 
$\frac{\pi}{2}$ phase-shift gate no state is left unchanged by the gate operation, which
explains why maximal fidelity is slightly decreased for small $\beta /\gamma$.
 
The stability of the adiabatic gates to decay of the excited state in the adiabatic limit is confirmed 
as the fidelities of the adiabatic gates tend to unity in the large $\Omega{T}$ limit. However, the 
fidelity of the revivals will no longer reach unity when there is decay of the excited state. In the 
adiabatic implementation of the $\frac{\pi}{2}$ phase-shift gate the maximal fidelity (red line) is 
identity since $|0\rangle$ is decoupled from the dynamics and therefore unaffected by the decay. For the Hadamard gate no state is decoupled from the dynamics and therefore the maximum fidelity is low for small $\Omega{T}$.

\subsubsection{Dephasing}
Dephasing is hard to eliminate in some implementations of holonomic gates, for example in 
superconducting Josephson junctions \cite{koch}. Therefore, we study the robustness of the 
non-adiabatic and adiabatic schemes to dephasing. More precisely, we consider dephasing 
in the $|0\rangle, |e\rangle$ and $|1\rangle, |e\rangle$ bases. The effect of dephasing 
is modeled by the Lindblad equation
\begin{eqnarray}\label{lind2}
\dot{\varrho}_t &=& -i[H(t),\varrho_t] \nonumber\\
&&+ \sum_{i=0,1}\left(2L_{ke} \varrho_t L_{ke}^{\dagger} - L_{ie}^{\dagger} L_{ie} \varrho_t -
\varrho_t L_{ie}^{\dagger} L_{ie}\right) ,
\end{eqnarray}
where $\varrho_t$ is the density operator, $L_{ke} = \sqrt{\epsilon}( \ket{e} \bra{e}- \ket{k} \bra{k})$ 
are the Lindblad operators, and $H(t)$ is either $H^{(\textrm{na})} (t)$ or $H^{(\textrm{a})} (t)$. 
We use the assumption that the time scale of the process underlying the dephasing is short 
compared to the time scale of the pulses (Markovian approximation). We again compare the 
resulting fidelities of the non-adiabatic implementations of the two test gates with those of  
their corresponding adiabatic implementations. 

In the non-adiabatic case, the operation parameters are $\beta$ and the total time $t_{r}$ 
between preparation and read-out. The two dimensionless parameters describing the dynamics 
are $\frac{\beta}{\epsilon}$ and $\epsilon{t_{r}}$. Since the qubit space is not a decoherence 
free subspace of the dephasing, the full time between preparation and read-out is relevant 
for the fidelity of the gate operation. The best fidelity is thus achieved when the total time is 
reduced to only the time required to implement the gates. To study the effect of dephasing 
on gate operation we have therefore assumed that gates are implemented immediately following 
preparation and that read-out is made immediately afterwards. This assumption reduces the 
relevant time to $\tau\propto\frac{1}{\beta}$ and the only relevant dimensionless parameter 
is therefore $\frac{\beta}{\epsilon}$. 

The relevant operation parameters in the adiabatic case are the coupling strength $\Omega$ 
and run-time $T$, and the dimensionless parameters describing the dynamics can be 
chosen as $\Omega{T}$ and $\epsilon T$. In the simulations we consider both a fixed coupling 
strength $\Omega_{0}$ and vary the run-time, and a fixed run-time $T_{0}$ and vary the coupling strength.

\begin{figure*}[htb!]
\centering
\includegraphics[width=0.325\textwidth]{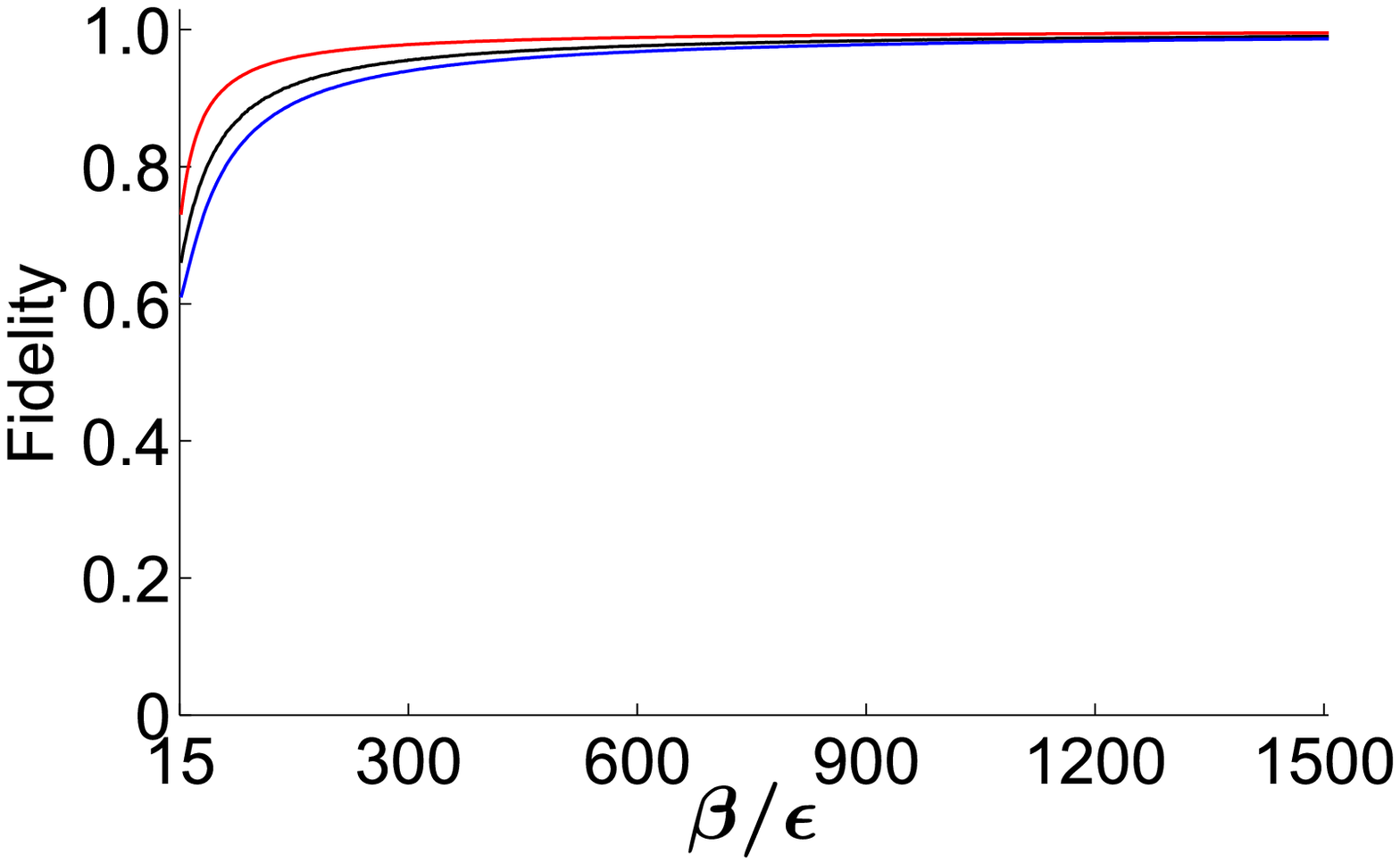}\phantom{.}
\includegraphics[width=0.32\textwidth]{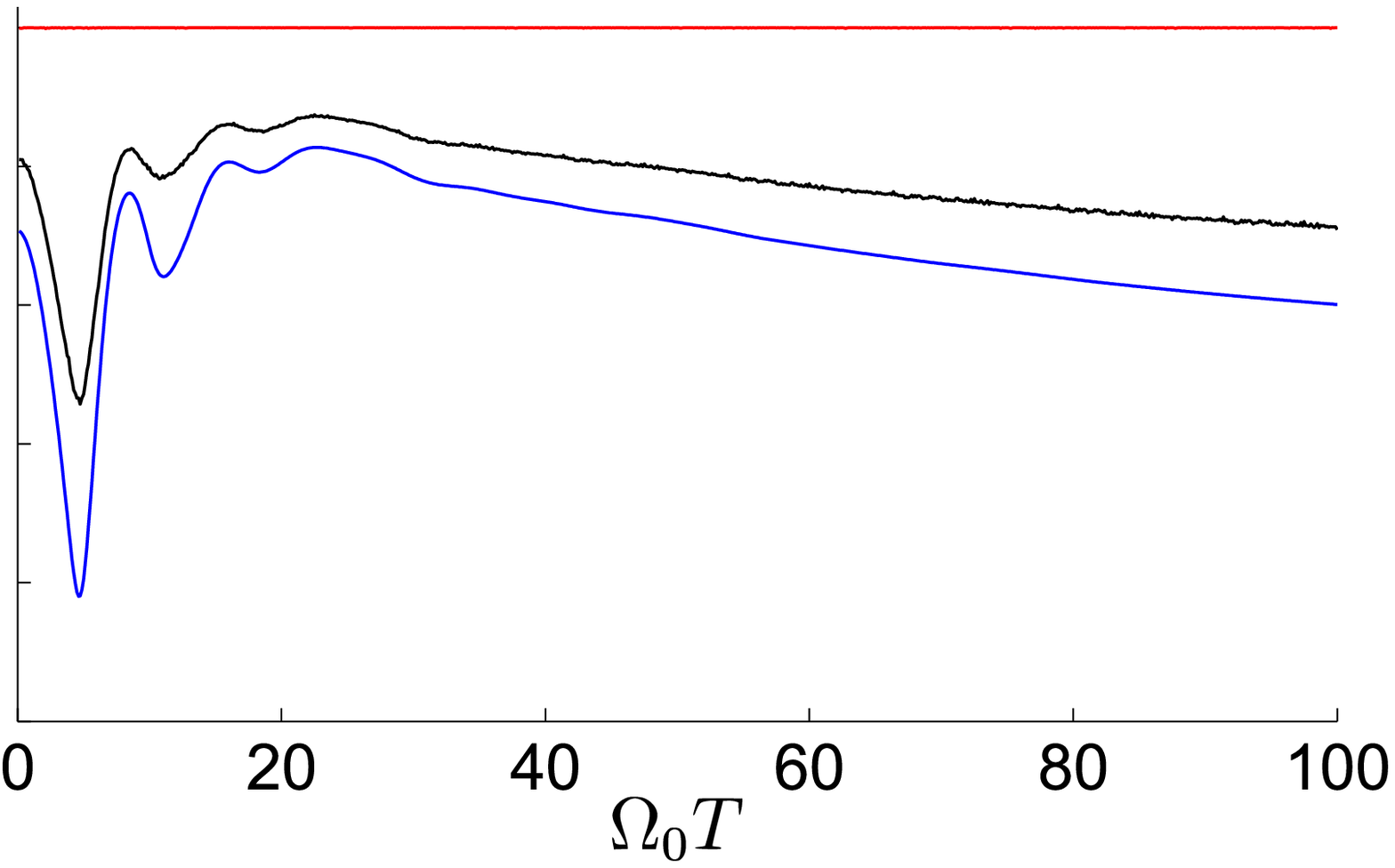}\phantom{.}
\includegraphics[width=0.32\textwidth]{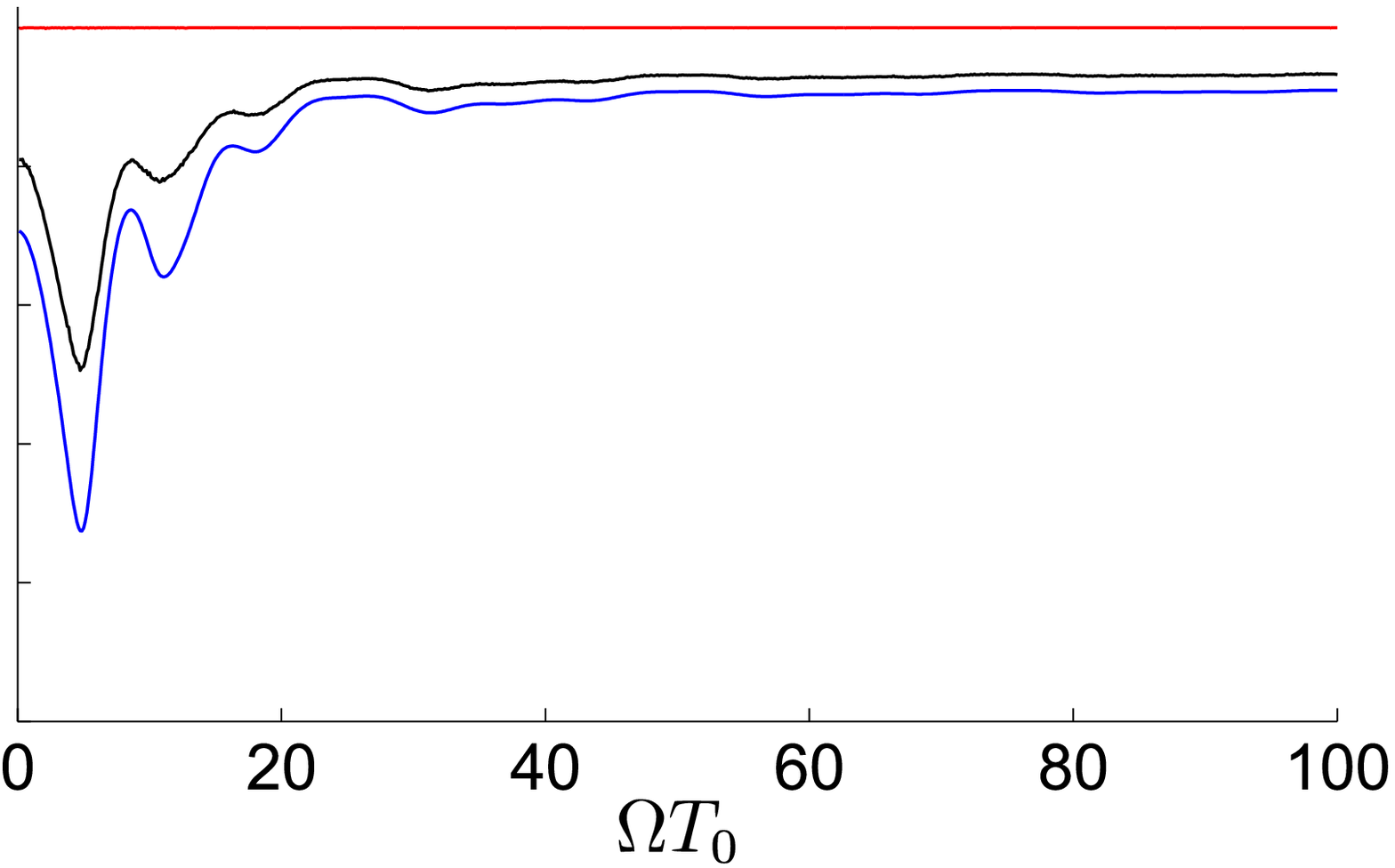}\phantom{.}
\centering
\includegraphics[width=0.325\textwidth]{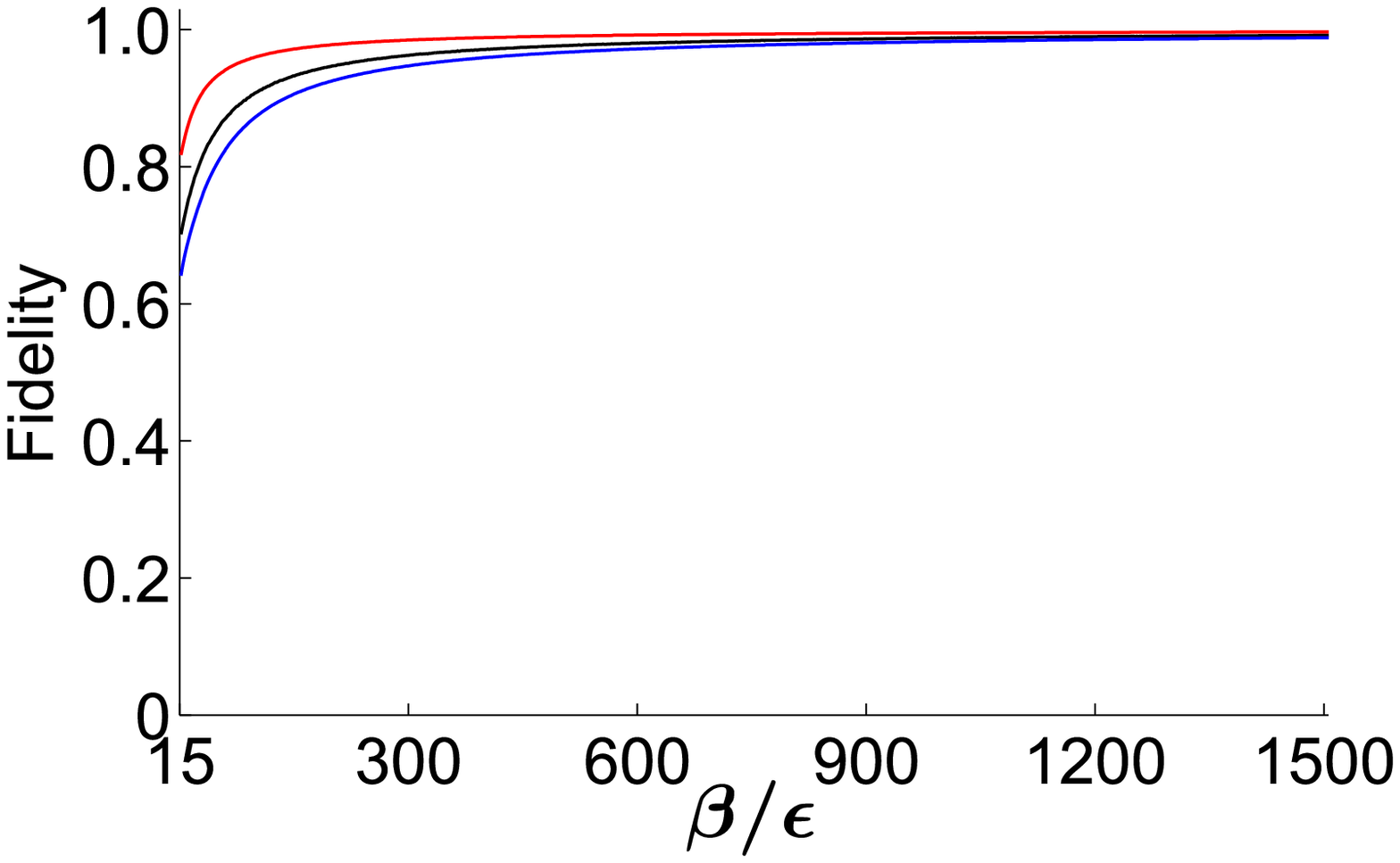}\phantom{.}
\includegraphics[width=0.32\textwidth]{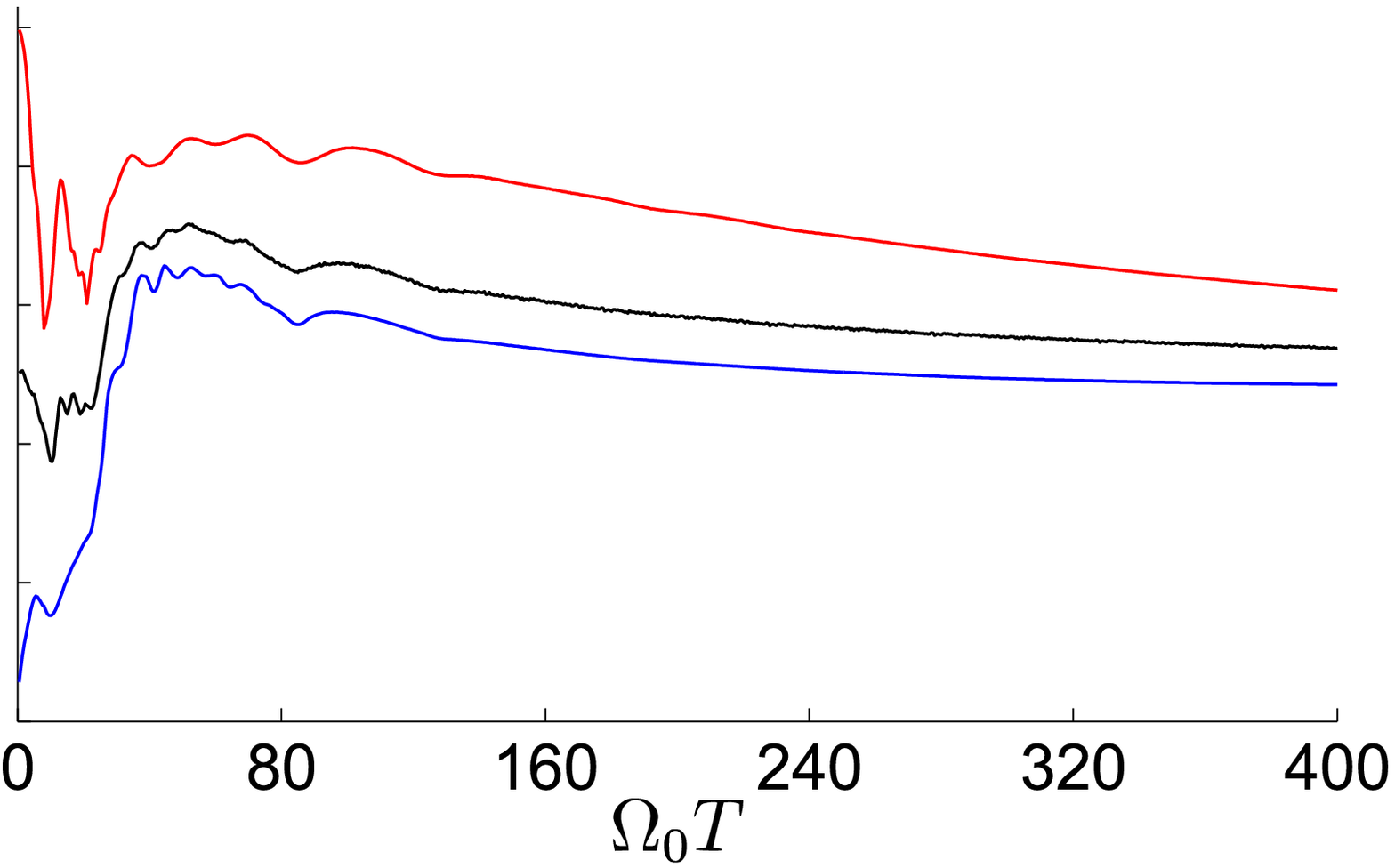}\phantom{.}
\includegraphics[width=0.32\textwidth]{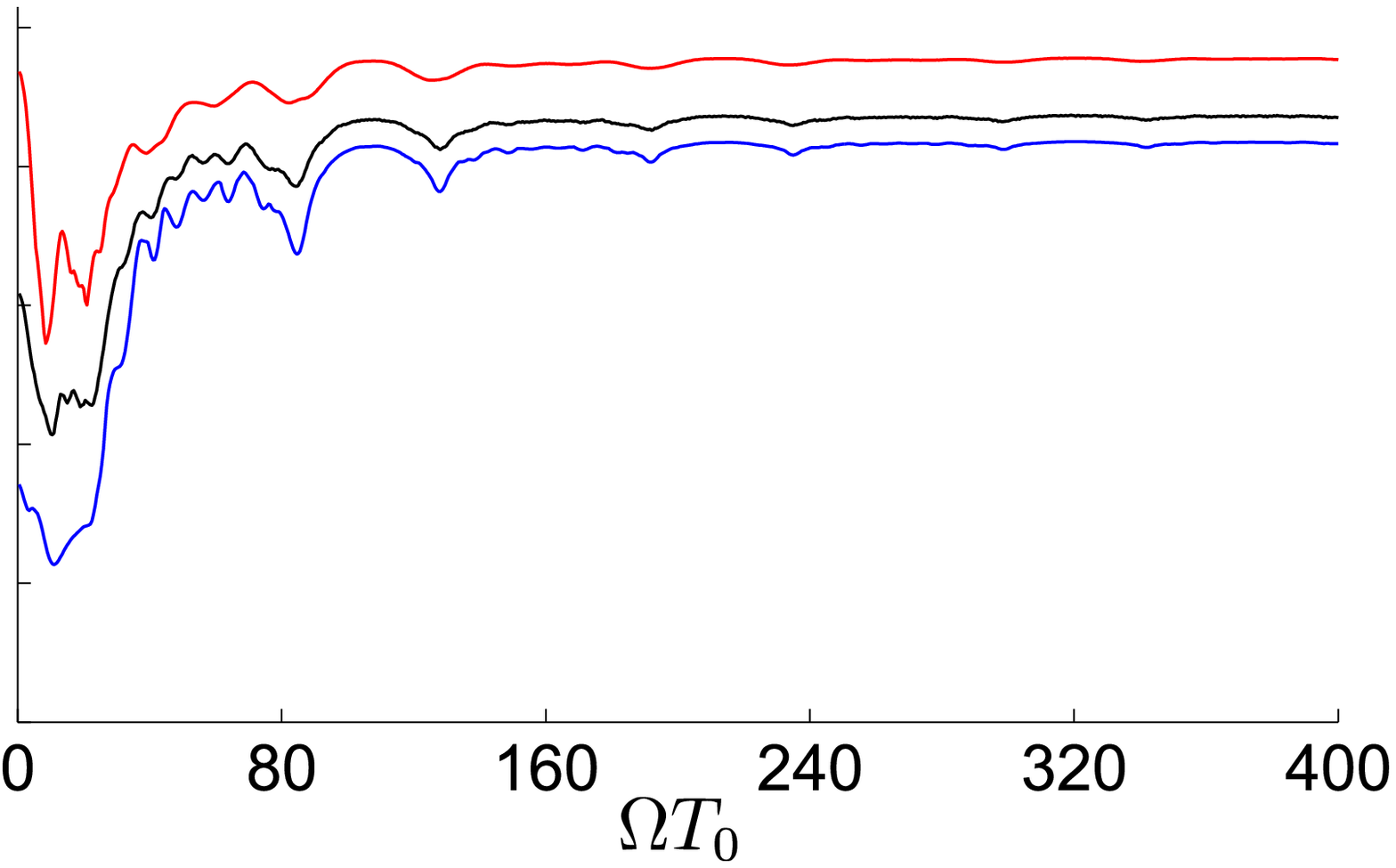}\phantom{.}
\caption{(Color online) Influence of dephasing in the $|0\rangle , |e\rangle$ and 
$|1\rangle , |e\rangle$ bases, on the non-adiabatic and adiabatic holonomic 
$\frac{\pi}{2}$ phase-shift gate (upper) and Hadamard gate (lower). The effect is quantified from top to bottom in 
terms of maximum (red), average (black), and minimum (blue) fidelities. The three panels 
show from left to right, the non-adiabatic gate with dephasing (left) and the adiabatic gate
with dephasing, for a fixed coupling strength (middle) and fixed run-time (right).
Choosing hyperbolic secant $\pi$-pulses with amplitude $\beta$, the non-adiabatic fidelities 
are plotted as functions of the dimensionless quantity $\beta /\epsilon$. We plot the adiabatic fidelities as functions of the dimensionless quantities $\Omega_{0} T$ and $\Omega T_{0}$, where 
$\Omega$ is the time independent global strength of field couplings, $T$ is the run-time 
of the gate, and $\Omega_{0}$ and $T_{0}$ are particular fixed values of these quantities. For the 
adiabatic gates, we have chosen $\Omega_{0} /\epsilon = 78.125$ for the case with fixed coupling 
strength and for the case of fixed run-time we have chosen $\epsilon{T_{0}}=0.16$ for the 
$\frac{\pi}{2}$ phase-shift gate and $\epsilon{T_{0}}=0.32$ for the Hadamard gate. In the 
non-adiabatic case, we have chosen $\epsilon t_{s} = 0.128$, where $t_{s}$ is the temporal 
separation of the two pulses. $t_{s}$ is chosen sufficiently large to avoid pulse overlap for 
the $\beta /\epsilon$ range shown in the left panels.}
\label{fig:graphs4}
\end{figure*}

In Fig. \ref{fig:graphs4}, we show the fidelities of the test gates, computed using Eq. (\ref{lind2}). 
The fidelities are plotted as functions of the dimensionless quantity $\beta/\epsilon$ in the non-adiabatic case, as well as $\Omega_{0} T$ and $\Omega T_{0}$ in the adiabatic case for fix coupling strength and fixed run-time, respectively.
For the adiabatic case we 
have chosen $\Omega_{0} /\epsilon = 78.125$, in the case of fixed coupling strength, 
and in the case of fixed run-time we have chosen $\epsilon{T_{0}}=0.16$ for the $\frac{\pi}{2}$ 
phase-shift gate and $\epsilon T_{0} = 0.32$ for the Hadamard gate. These choices are made to 
make the effect of dephasing non-negligible. Furthermore, in the non-adiabatic case we have 
chosen $\epsilon t_{s} = 0.128$, which guarantees that there is no pulse overlap for the 
$\beta /\epsilon$ range shown.

The fidelities of the non-adiabatic gates tend monotonically to unity in the large 
$\beta /\epsilon$ limit for both test gates (left panels). Thus, the non-adiabatic version of 
the holonomic $\frac{\pi}{2}$ phase-shift gate can be made resilient to the dephasing channel 
by employing sufficiently short pulses and reducing idle time before and after the pulses. 

The adiabatic gates are not stable to dephasing in the limit $T\to\infty$ and we see a gradual 
decline in the fidelities as run-time increases.  If the run-time is fixed and $\Omega$ 
is increased the fidelities stabilize at some value below unity that is a function of the parameter 
$\epsilon{T_{0}}$. To have high fidelity in the adiabatic case it is necessary that $\Omega$ can be 
made large compared to $\epsilon$ so that the adiabatic approximation is valid for a 
run-time at which the effect of dephasing is still negligible. Alternatively, $\Omega$ should be large 
enough so that the effect of dephasing is negligible at some $T$ corresponding to a revival of the 
fidelity.
 
\subsection{Parametric control}
\label{lb}
\subsubsection{Detuning}
\label{1bb}
\begin{figure*}[htb!]
\centering
\includegraphics[width=0.325\textwidth]{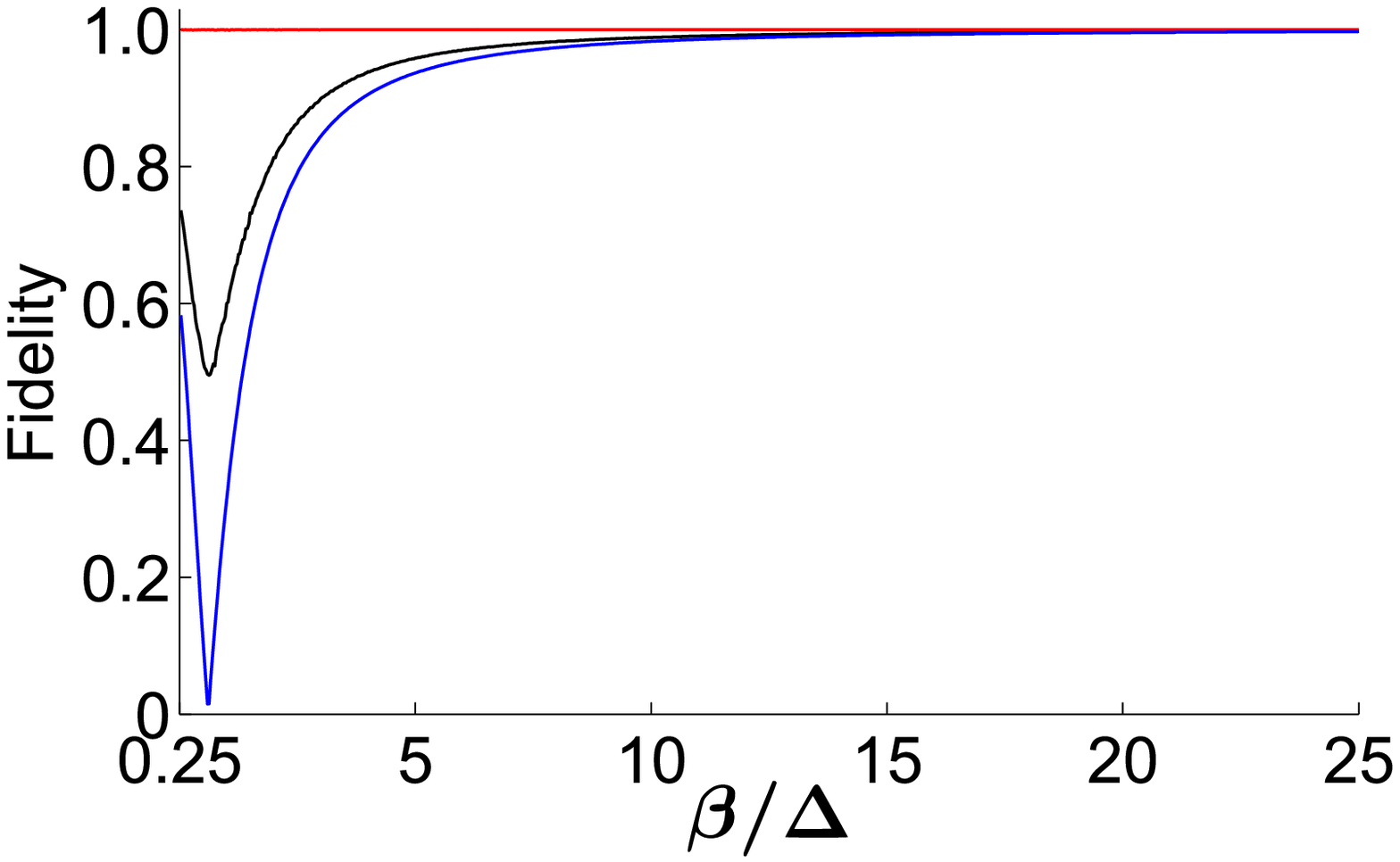}\phantom{.}
\includegraphics[width=0.32\textwidth]{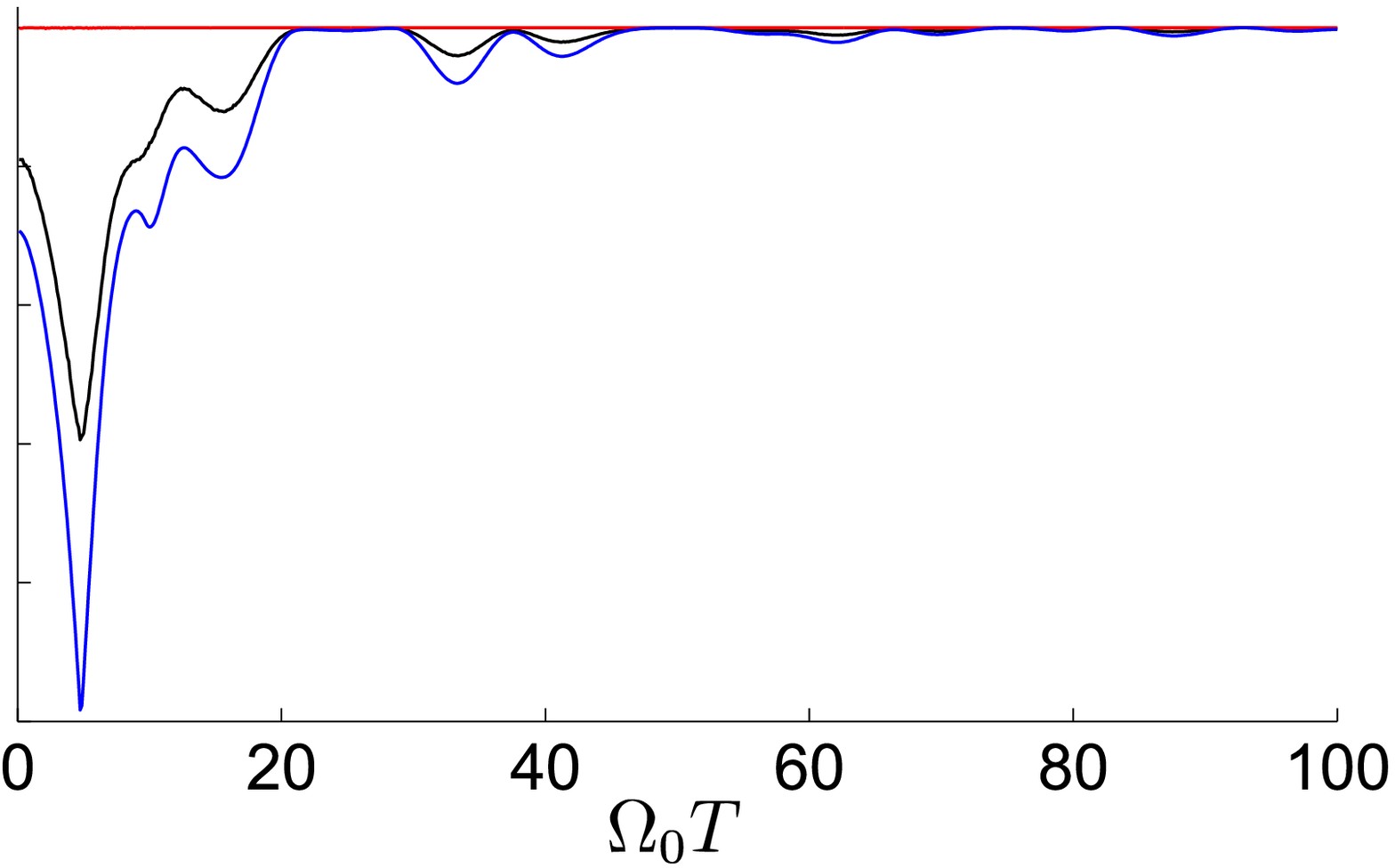}\phantom{.}
\includegraphics[width=0.32\textwidth]{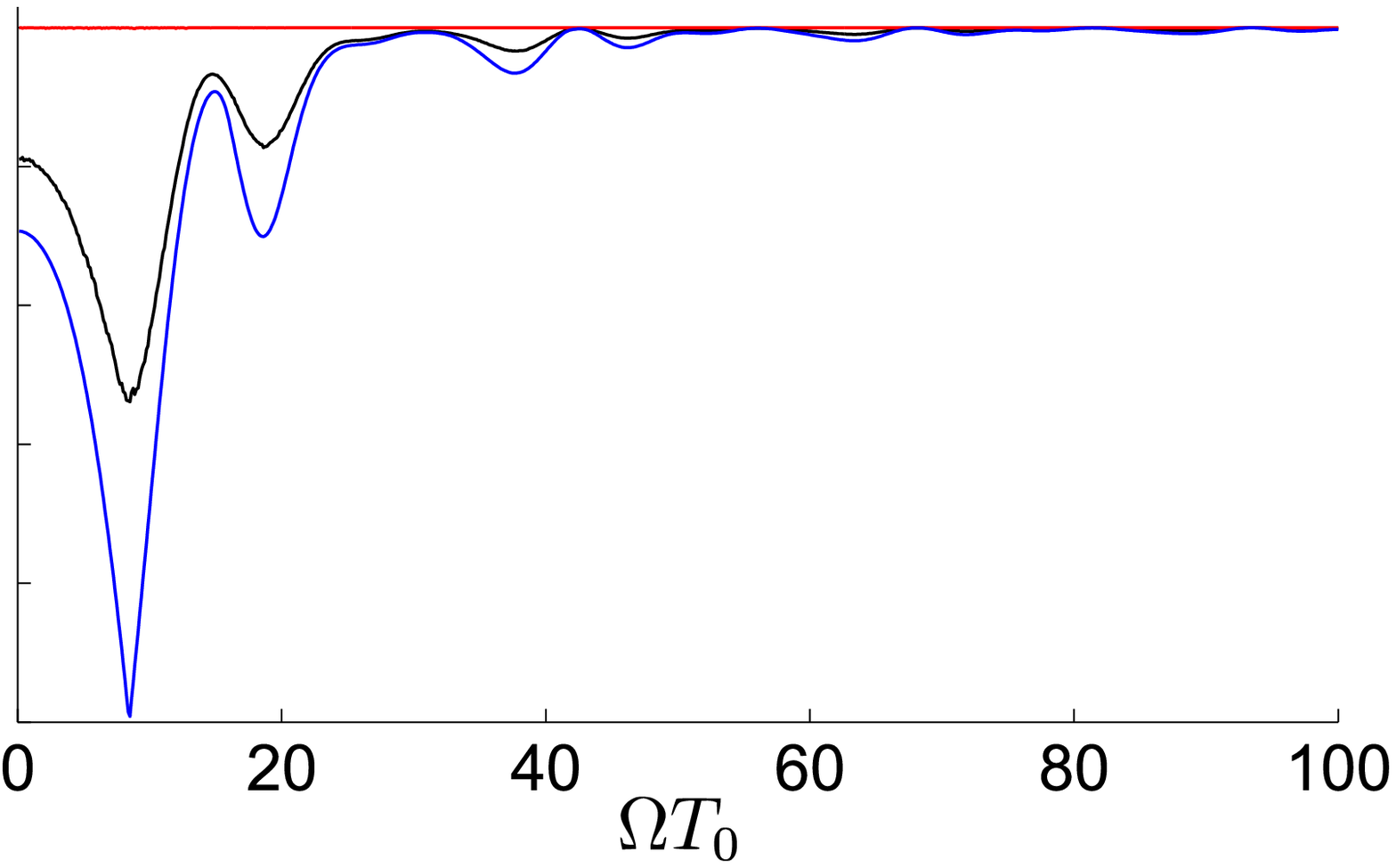}\phantom{.}
\centering
\includegraphics[width=0.325\textwidth]{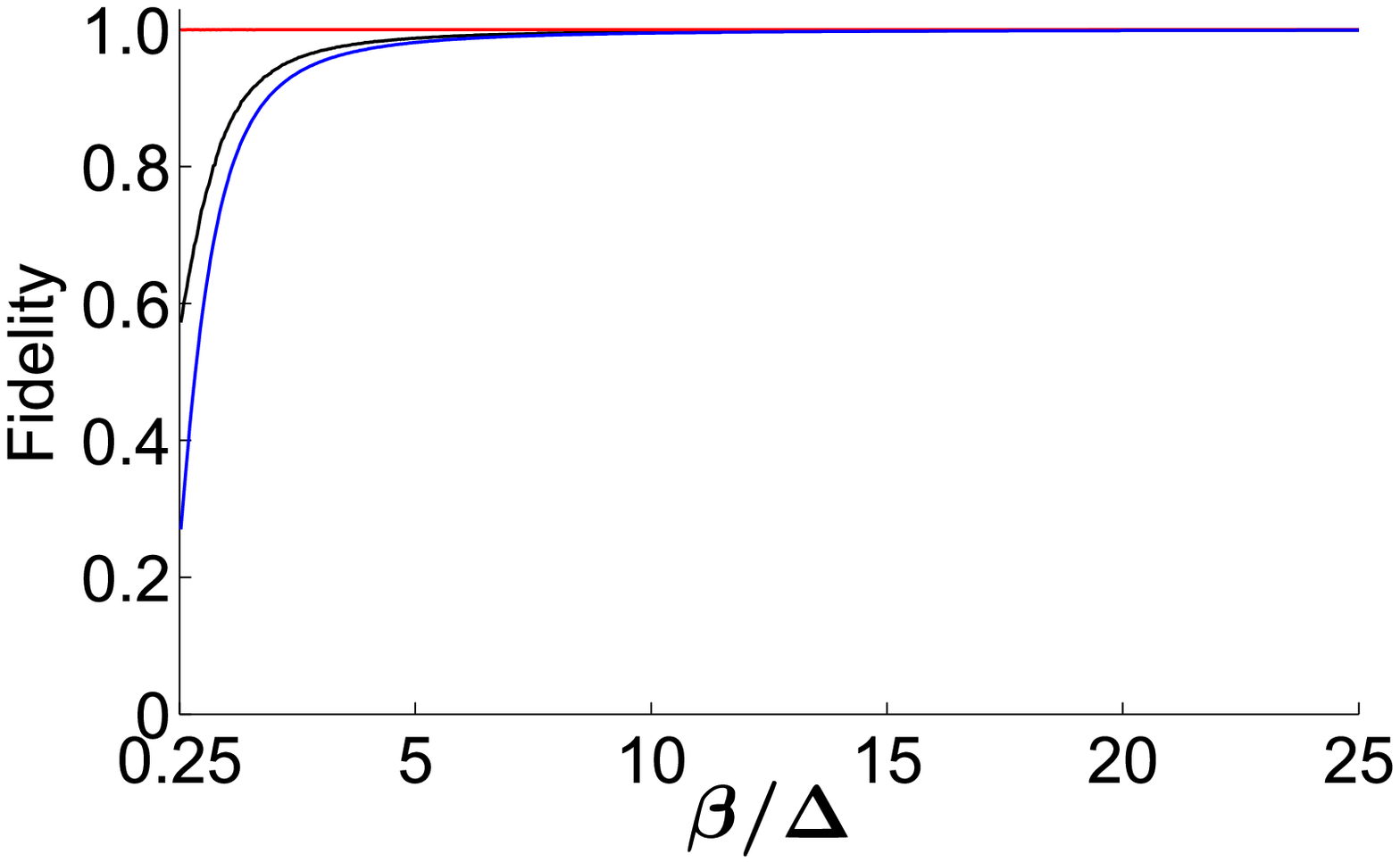}\phantom{.}
\includegraphics[width=0.32\textwidth]{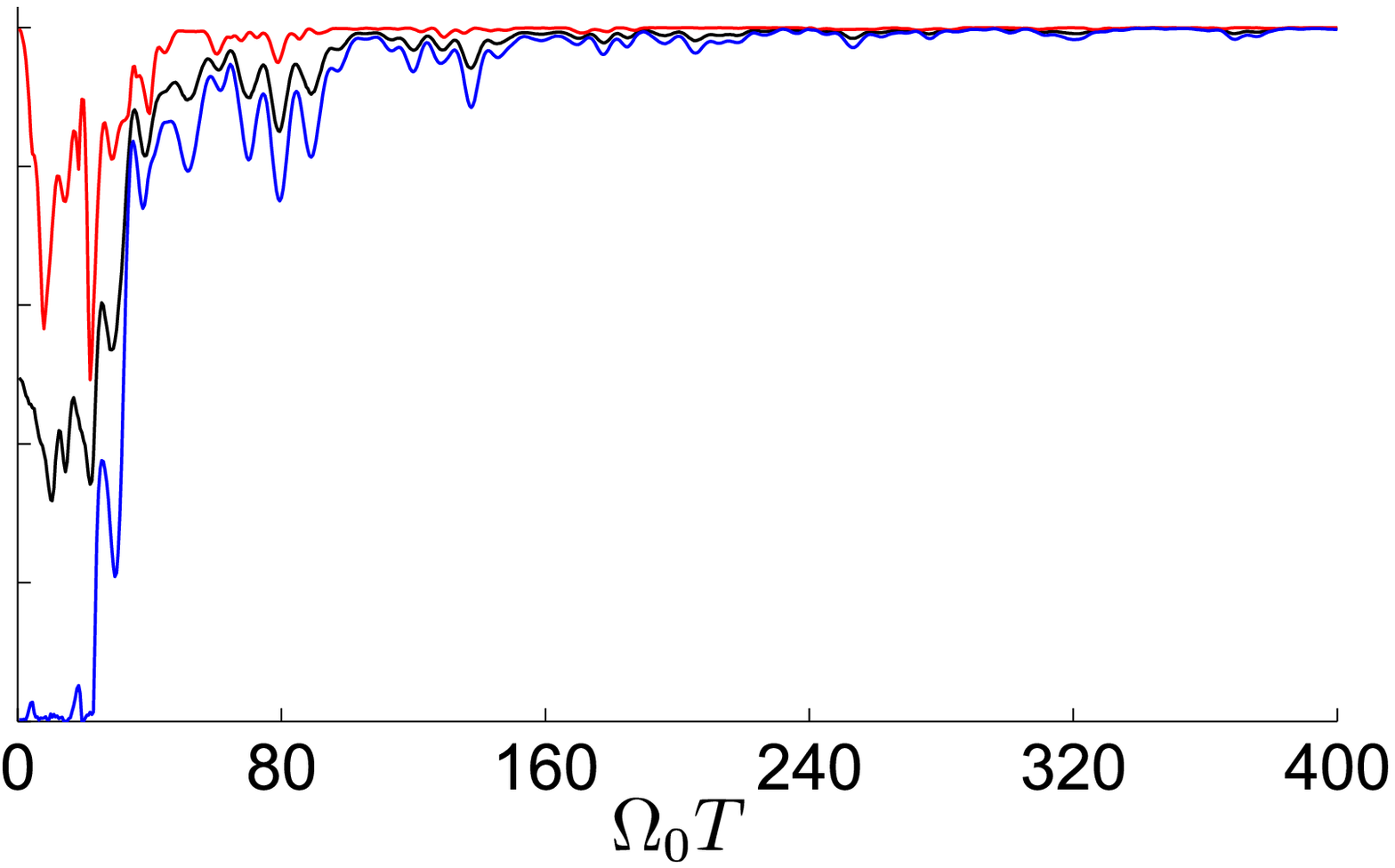}\phantom{.}
\includegraphics[width=0.32\textwidth]{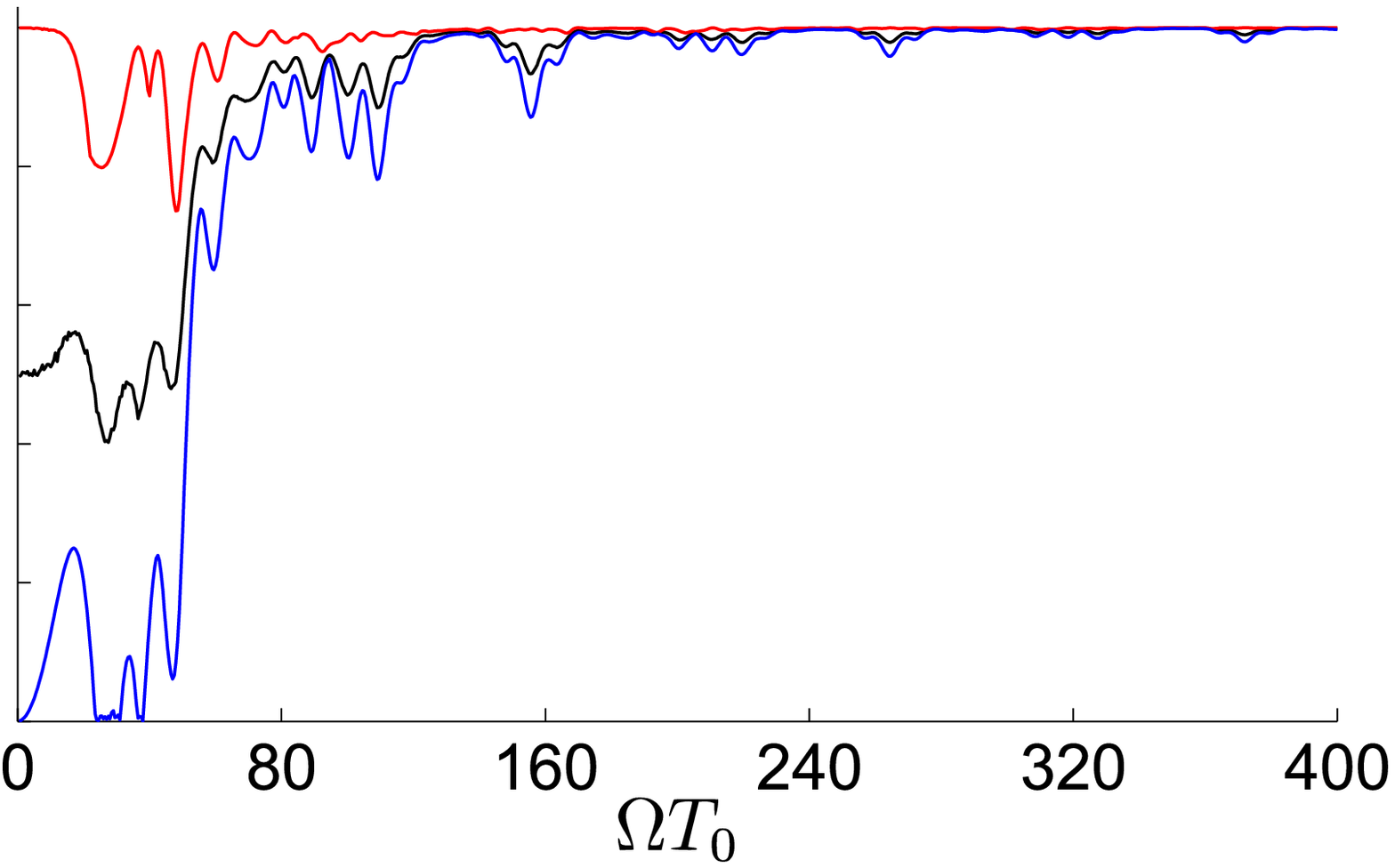}\phantom{.}
\caption{(Color online) Influence of a constant mean detuning $\Delta$ and zero relative 
detuning on the non-adiabatic and adiabatic holonomic $\frac{\pi}{2}$ phase-shift gate 
(upper) and Hadamard gate (lower). The effect is quantified from top to bottom in 
terms of maximum (red), average (black), and minimum (blue) fidelities. The three panels show from left to right, the 
non-adiabatic gate with mean detuning (left) and the adiabatic gate with mean detuning, 
for a fixed coupling strength (middle) and fixed run-time (right). Choosing hyperbolic 
secant $\pi$-pulses with amplitude $\beta$, the non-adiabatic fidelities are plotted as 
functions of the dimensionless quantity $\beta /\Delta$. We plot the adiabatic fidelities as functions of the dimensionless quantities $\Omega_{0} T$ and $\Omega T_{0}$, where 
$\Omega$ is the time independent global strength of field couplings, $T$ is the run-time 
of the gate, and $\Omega_{0}$ and $T_{0}$ are particular fixed values of these quantities. For the adiabatic gates 
we have chosen $\Omega_{0} /\Delta_{01a} = 6.25$ for the case with fixed coupling strength and 
for the case of fixed run-time we have chosen $\Delta_{01a}{T_{0}}=16$ for the $\frac{\pi}{2}$ 
phase-shift gate and $\Delta_{01a}{T_{0}}=64$ for the Hadamard gate. In the non-adiabatic case we 
have chosen $\Delta t_{s} = 80$, where $t_{s}$ is the temporal separation of the two pulses.
$t_{s}$ is chosen sufficiently large to avoid pulse overlap for the $\beta /\Delta$
range shown in the left panels.}
\label{fig:graphs2}
\end{figure*}

We assume that the $\ket{j}\leftrightarrow \ket{e}$ transition is driven by a laser pulse with frequency 
$\nu_j$. The associated detuning is $\Delta_j = 2\pi \nu_j - \omega_{je}$, where $\omega_{je}$ 
is the corresponding energy spacing. Ideally, the detunings must vanish in the non-adiabatic 
case while they must be all equal in the adiabatic setting. Here, we examine the effect of deviations 
from these ideal values on the gate fidelity. To simplify the analysis, we limit the study to time 
independent detunings. 

Non-zero detunings give rise to additional diagonal terms in the Hamiltonian. In the 
non-adiabatic case, we have 
\begin{eqnarray}
\label{punk}
H_{\Delta}^{(\textrm{na})} (t) & = &  \Delta_{0}\ket{0} \bra{0}+\Delta_{1}\ket{1} \bra{1} + 
H^{(\textrm{na})} (t) , 
\end{eqnarray}
where $H^{(\textrm{na})} (t)$ is the ideal Hamiltonian in Eq. (\ref{eq:ideal_na}). Similarly, in the 
adiabatic case, we have  
\begin{eqnarray} 
\label{punk2} 
H_{\Delta}^{(\textrm{a})} (t) & = & \Delta_{0}\ket{0} \bra{0} + \Delta_{1}\ket{1} \bra{1} + 
\Delta_{a}\ket{a} \bra{a} + H^{(\textrm{a})} (t) , 
\nonumber\\
\end{eqnarray}
where now $H^{(\textrm{a})} (t)$ is the ideal Hamiltonian in Eq. (\ref{eq:ideal_a}). If two detunings 
are equal but different from the third, there will only be one dark state, and if all three detunings are different 
from each other there will be no dark state at all. 
Thus, deviations from the $\Delta_{0}=\Delta_{1}=\Delta_{a}$ constraint destroy the dark state 
structure and the associated 
holonomy. 

Note that we express $H_{\Delta}^{(\textrm{na})} (t)$ and $H_{\Delta}^{(\textrm{a})} (t)$ in frames 
that are ``co-rotating'' with their respective detuned driving fields. Therefore, to compare the 
output of the gate operation $\varrho_{\textrm{out}}(t)$ generated by the detuned Hamiltonians, with the 
output of the desired gate operation $U(C)$, we must transform $\varrho_{\textrm{out}}(t)$ to the 
frame co-rotating with the ideal driving fields. This is done through the transformation 
$\varrho_{\textrm{out}}(t)\to{e}^{iSt}\varrho_{\textrm{out}}(t)e^{-iSt}$, where $S=\Delta_{0}|0\rangle\langle{0}|+\Delta_{1}|1\rangle\langle{1}|$ in the non-adiabatic case and $S=\Delta_{0}|0\rangle\langle{0}|+\Delta_{1}|1\rangle\langle{1}|+\Delta_{a}|a\rangle\langle{a}|$ in the adiabatic case.
Note also that these frame rotations will remove the effect of the diagonal terms in Eqs. (\ref{punk}) and (\ref{punk2}) for the part of the evolution where the driving fields vanish.

We study two principal cases. First, all detunings are set equal, and secondly, some of the detunings 
are assumed to be different. These cases are naturally captured by the mean and relative detunings. 
In the non-adiabatic setting, these read $\Delta=\frac{\Delta_{0}+\Delta_{1}}{2}$ and 
$\delta=\frac{\Delta_{0}-\Delta_{1}}{2}$, respectively. In the adiabatic case, we have the mean 
detuning $\Delta_{01a} = \frac{\Delta_{0} + \Delta_{1} + \Delta_{a}}{3}$ and two independent 
relative detunings $\delta_{01}=\frac{\Delta_{0}-\Delta_{1}}{2}$ and $\delta_{0a} =  
\frac{\Delta_{0}-\Delta_{a}}{2}$. In some implementations it may be easier to control the 
relative detuning of the driving fields, than the mean detuning.

\begin{figure*}[htb!]
\centering
\includegraphics[width=0.325\textwidth]{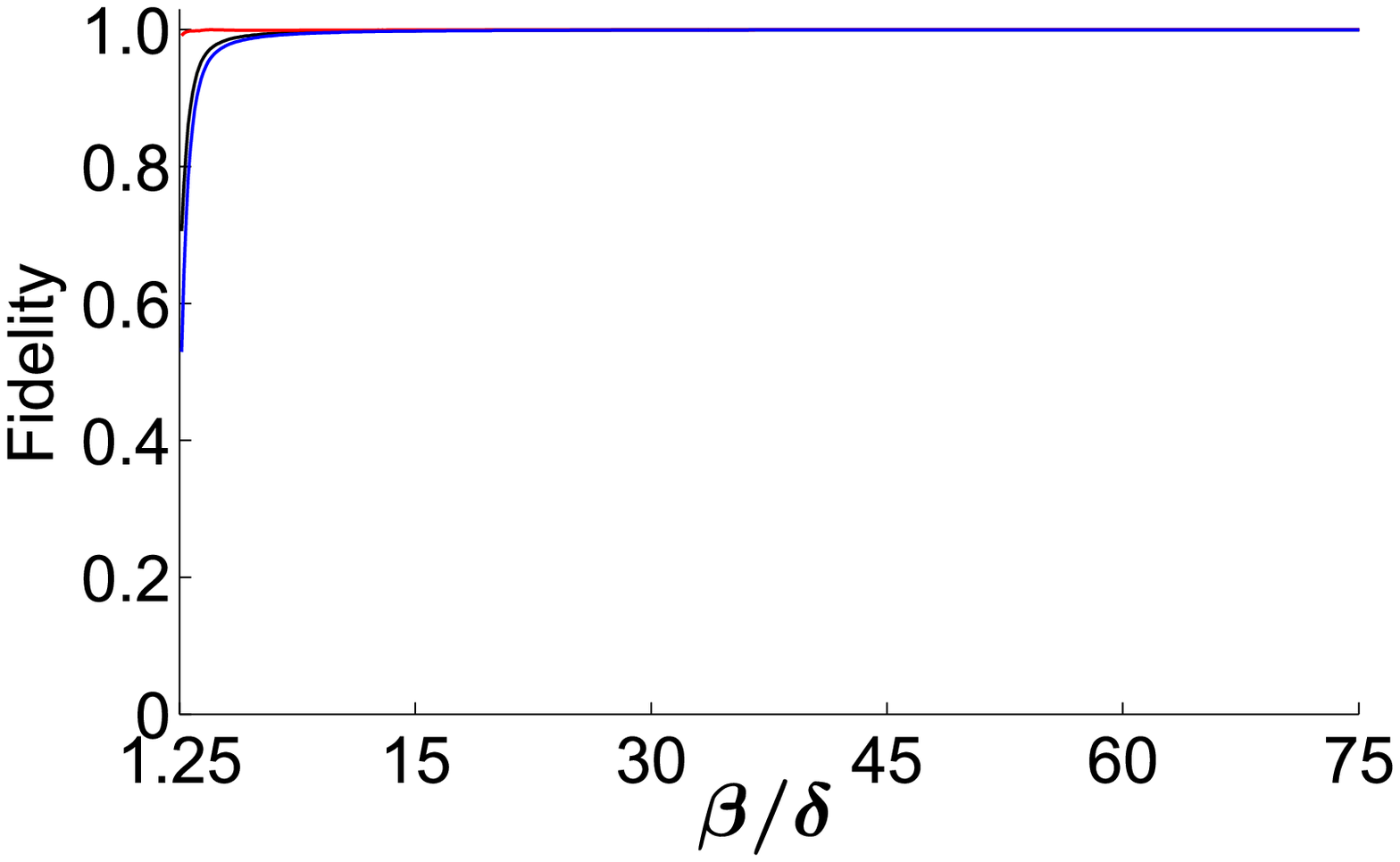}\phantom{.}
\includegraphics[width=0.32\textwidth]{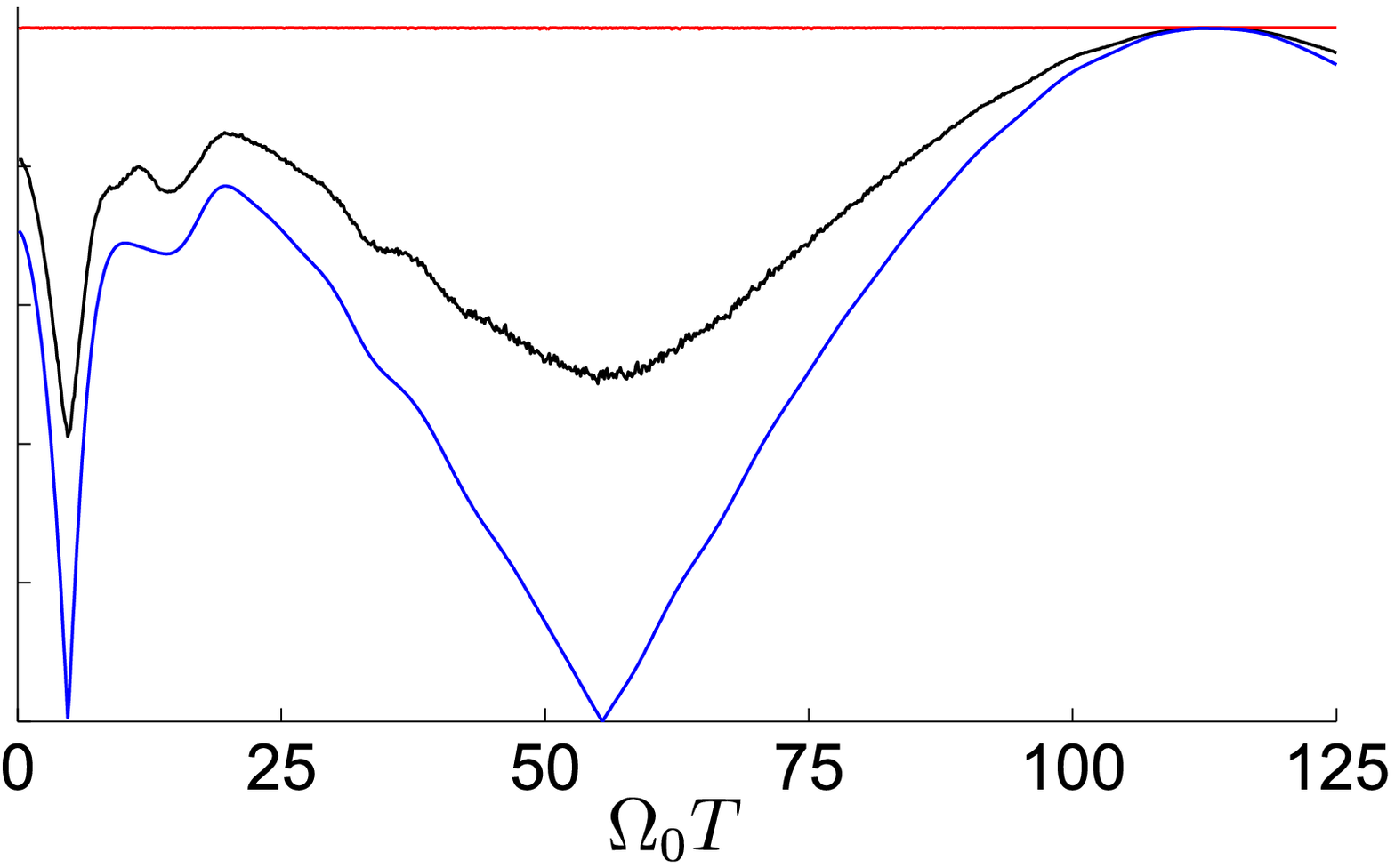}\phantom{.}
\includegraphics[width=0.32\textwidth]{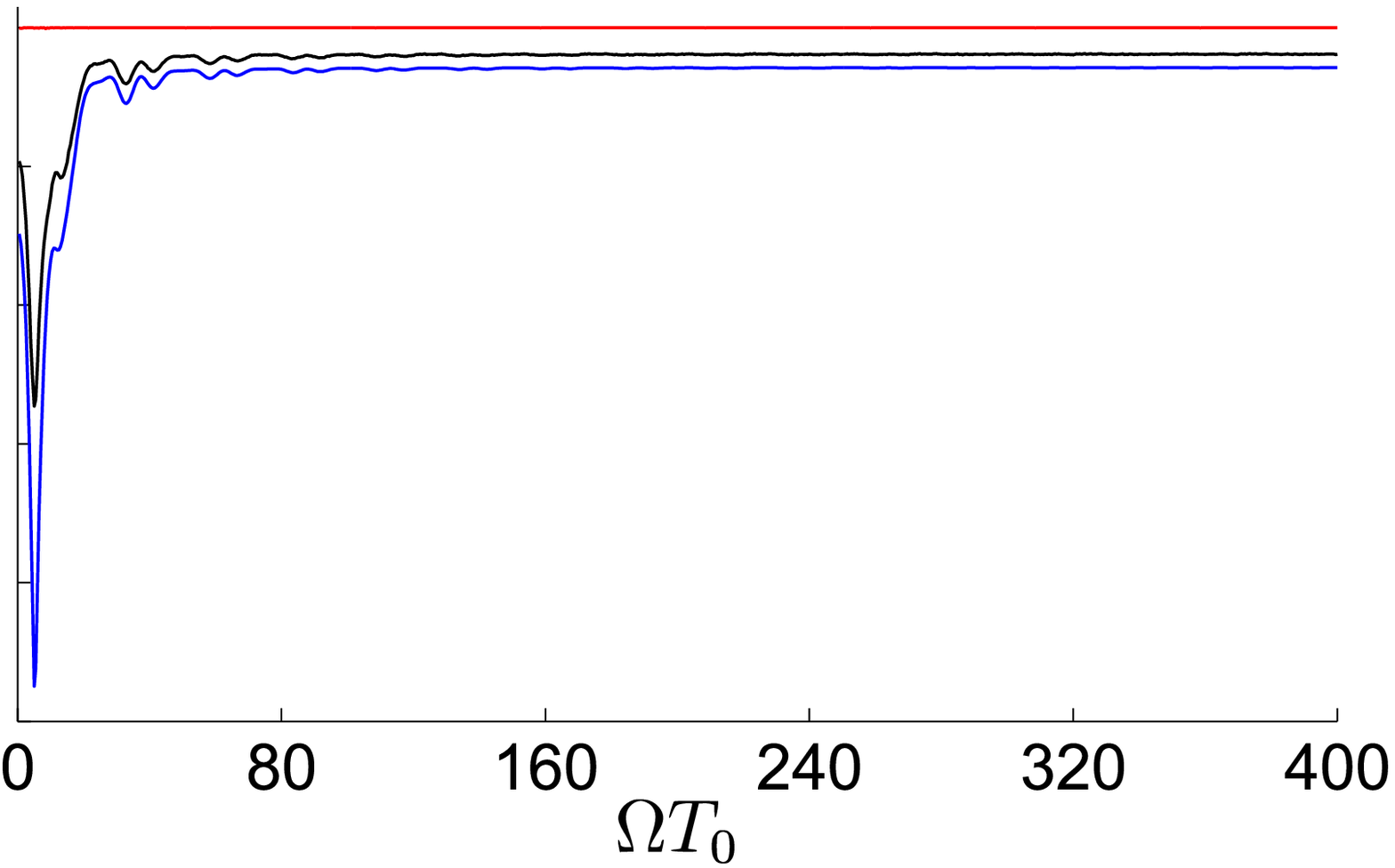}\phantom{.}
\centering
\includegraphics[width=0.325\textwidth]{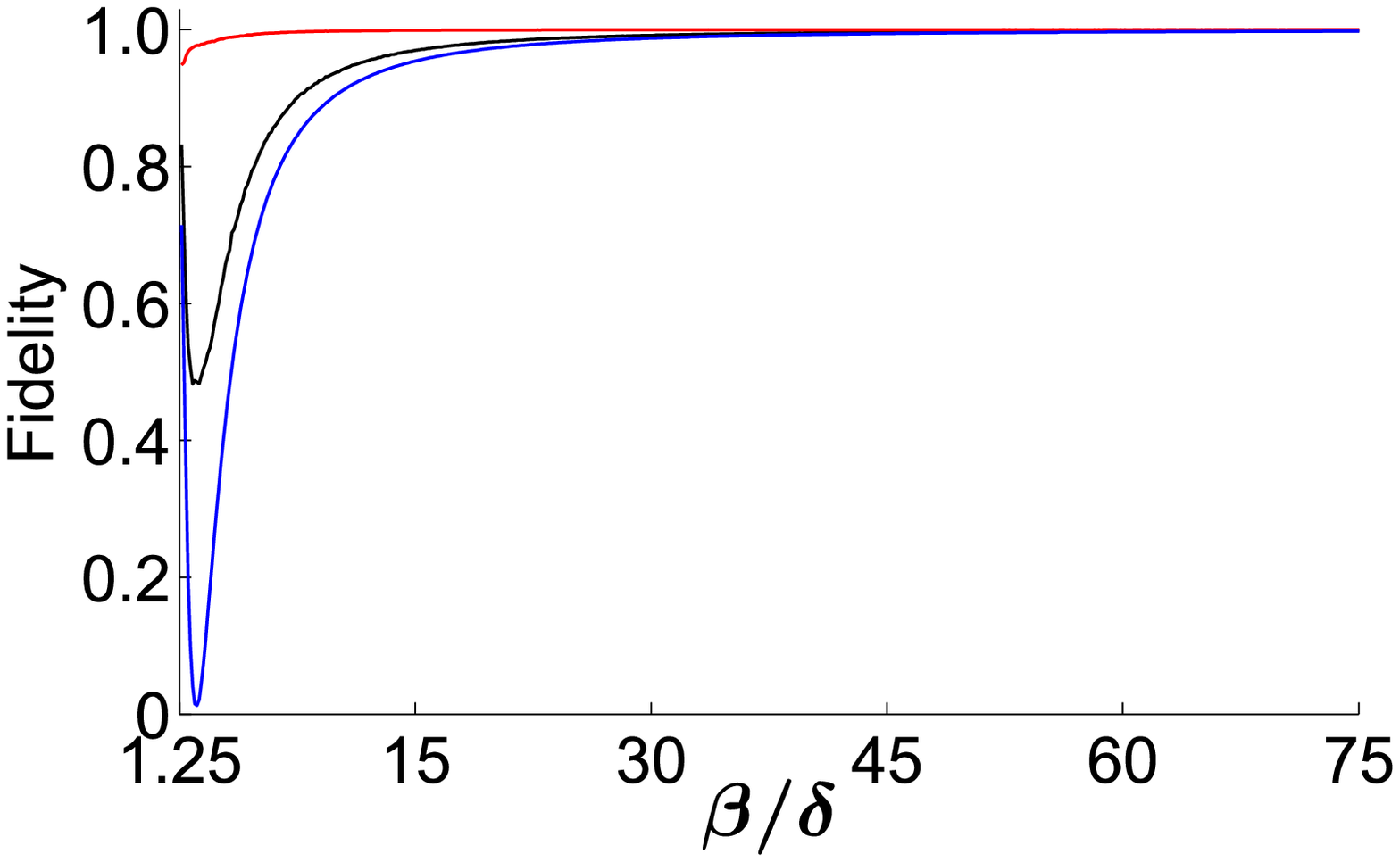}\phantom{.}
\includegraphics[width=0.32\textwidth]{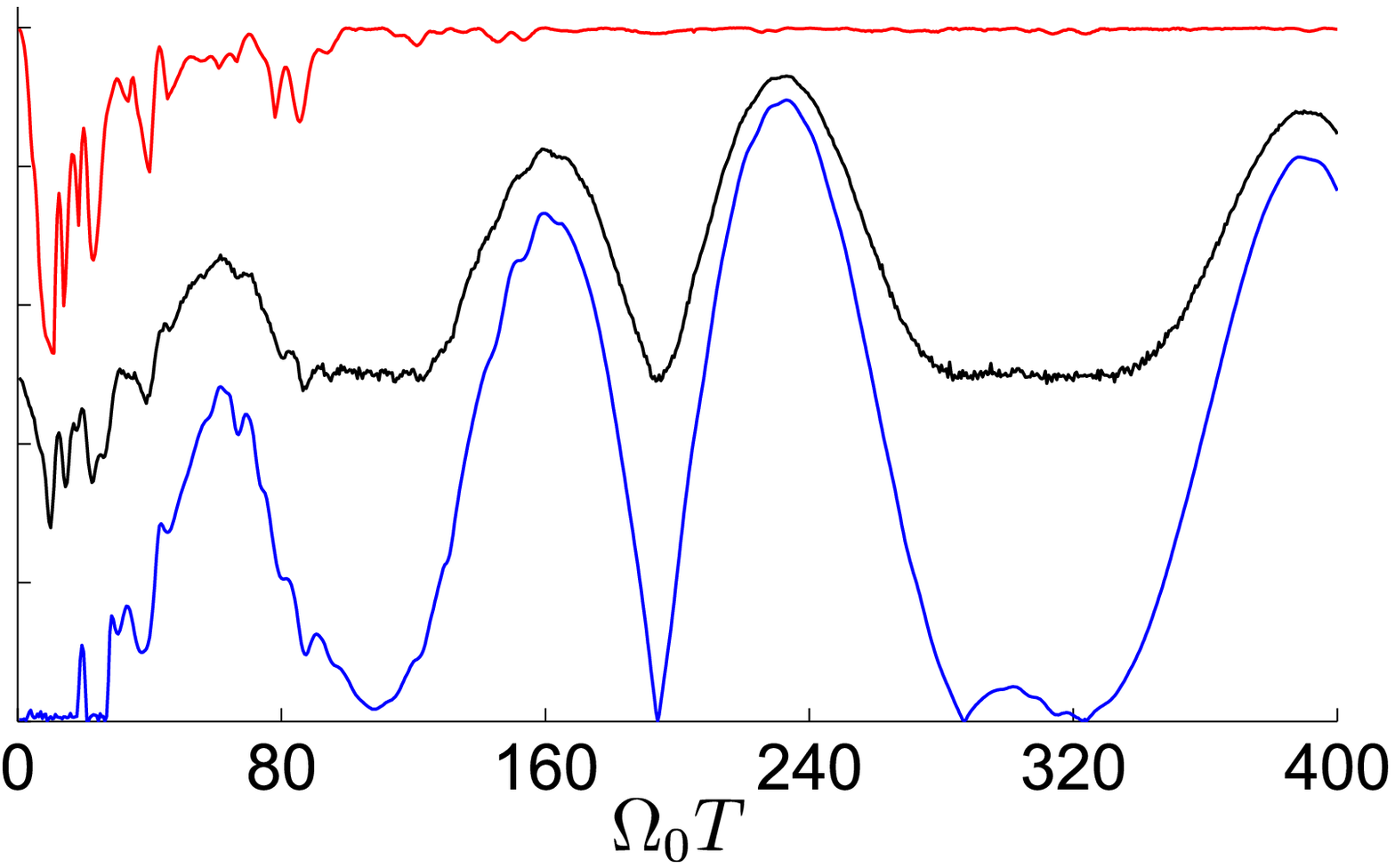}\phantom{.}
\includegraphics[width=0.32\textwidth]{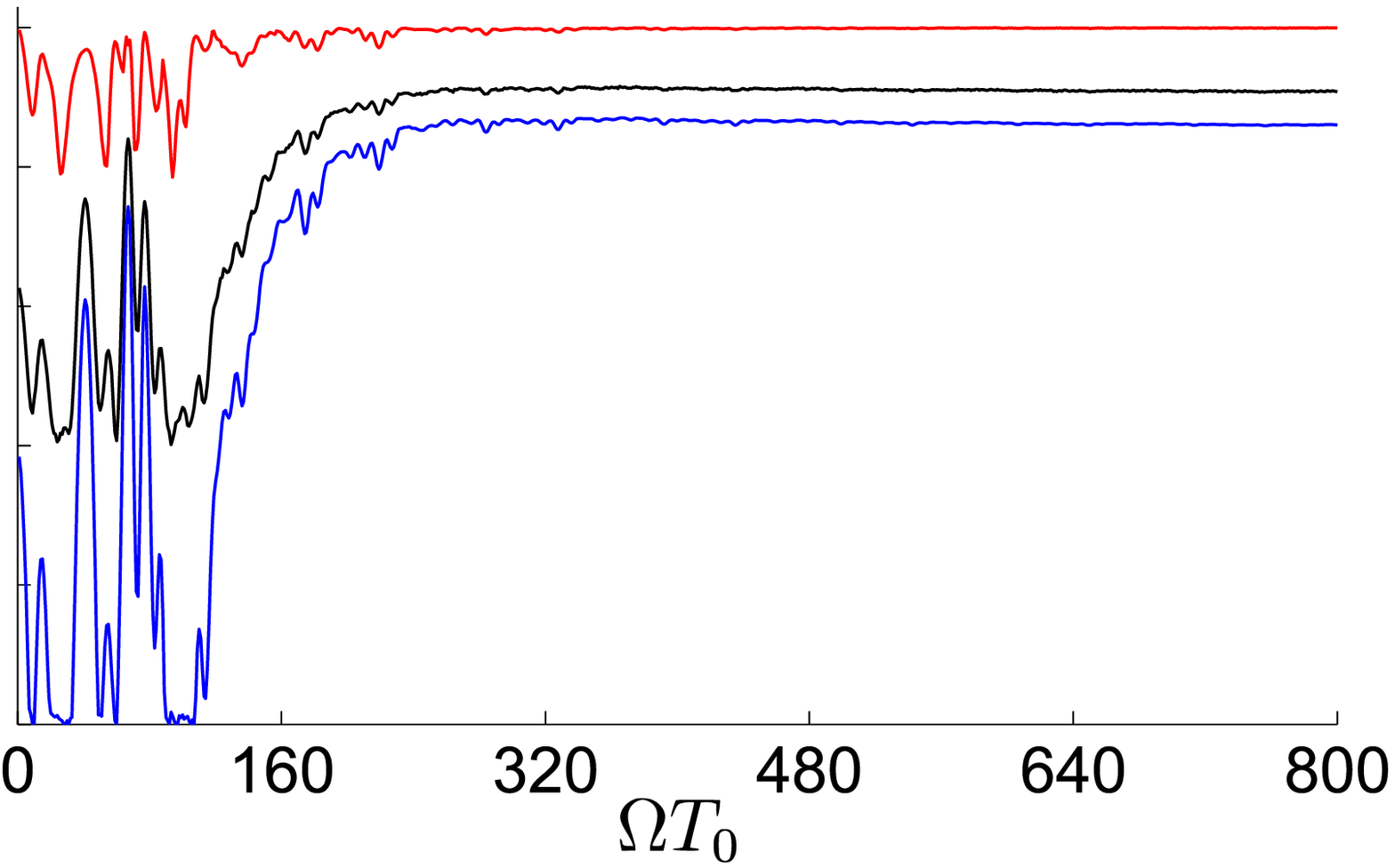}\phantom{.}
\caption{(Color online) Influence of a constant relative detuning $\delta$ on the non-adiabatic
and adiabatic holonomic $\frac{\pi}{2}$ phase-shift gate (upper) and Hadamard gate (lower).
The effect is quantified from top to bottom in 
terms of maximum (red), average (black), and minimum (blue) fidelities.
The three panels show, from left to right, the non-adiabatic gate with relative detuning and the 
adiabatic gate with relative detuning, for a fixed coupling strength (middle) and fixed run-time (right). Choosing hyperbolic secant $\pi$-pulses with amplitude $\beta$, the non-adiabatic 
fidelities are plotted as functions of the dimensionless quantity $\beta /\delta$.  We plot the adiabatic fidelities as functions of the dimensionless quantities $\Omega_{0} T$ and $\Omega T_{0}$, where 
$\Omega$ is the time independent global strength of field couplings, $T$ is the run-time 
of the gate, and $\Omega_{0}$ and $T_{0}$ are particular fixed values of these quantities.
For the adiabatic gates we have chosen $\Omega_{0} /\delta = 6.25$ for the case with fixed coupling 
strength and for the case of fixed run-time we have chosen $\delta{T_{0}}=2$ for the $\frac{\pi}{2}$ 
phase-shift gate and $\delta{T_{0}}=64$ for the Hadamard gate. In the non-adiabatic case we have 
chosen $\delta{t}_{s} = 16$, where $t_{s}$ is the temporal separation of the two pulses.
$t_{s}$ is chosen sufficiently large to avoid pulse overlap for the $\beta /\delta$
range shown in the left panels.}
\label{fig:graphs3}
\end{figure*}

First, we investigate how gate operation is affected in the non-adiabatic case when a 
constant mean detuning is introduced and the relative detuning is zero. For comparison we also 
include the adiabatic implementations with the same mean detuning and all relative detunings zero.

The dynamics of the non-adiabatic gate with $\Delta \neq 0$ and $\delta = 0$ depends on 
two dimensionless parameters that can be chosen as $\beta/\Delta$ and $\Delta t_r$. 
However since the entire dynamics is generated by the driving fields the relevant time is the pulse length $\tau\propto{\frac{1}{\beta}}$. Thus, $\beta/\Delta$ is the only relevant dimensionless parameter.

The relevant operation parameters in the adiabatic case are the coupling strength $\Omega$ and 
run-time $T$, and the dimensionless parameters describing the dynamics are $\Omega{T}$ 
and $\Delta{T}$. In the simulations we consider both a fixed coupling strength $\Omega_{0}$ and vary 
the run-time and a fixed run-time $T_{0}$ and vary the coupling strength.

In Fig. \ref{fig:graphs2}, we show the fidelities of the gates computed using Eqs. (\ref{punk}) 
and (\ref{punk2}) in the non-adiabatic and adiabatic case, respectively. The fidelities are plotted as functions of the dimensionless quantity $\beta/\Delta$ in the non-adiabatic case, as well as  $\Omega_{0} T$ and $\Omega T_{0}$ in the adiabatic case for fix coupling strength and fixed run-time, respectively. We choose $\Omega_{0} /\Delta_{01a} = 6.25$ in the case of 
fixed coupling strength. In the case of fixed run-time, we take $\Delta_{01a}T_{0}=16$ for the 
$\frac{\pi}{2}$ phase-shift gate and $\Delta_{01a}T_{0}=64$ for the Hadamard gate.  Furthermore 
$\Delta t_{s}= 80$, which guarantees that there is no pulse overlap for the $\beta /\Delta$ range 
shown in the figure.

With a constant mean detuning and zero relative detuning, the fidelities of the non-adiabatic gates 
tend to unity in the large $\beta /\Delta$ limit (left panels). This demonstrates that the non-adiabatic 
versions of the holonomic test gates can be made robust to constant mean detuning by employing 
sufficiently short pulses. The adiabatic scheme is stable to constant mean detuning in the 
$\Omega{T}\to\infty$ limit.

Next, we consider nonzero relative detuning. The adiabatic implementation of the $\frac{\pi}{2}$ 
phase-shift gate does not involve any driving field coupled to the $\ket{0} \leftrightarrow \ket{e}$ 
transition. For this reason we consider the case where the only nonzero detuning is 
$\delta\equiv{\Delta_{1}}$ of the driving field associated with the $\ket{1} \leftrightarrow \ket{e}$ 
transition.

The relevant dimensionless 
dynamical parameter for the non-adiabatic gates is $\beta/\delta$. The relevant operation parameters 
in the adiabatic case are the coupling strength $\Omega$ and run-time $T$, and the dimensionless 
parameters describing the dynamics are $\Omega{T}$ and $\delta{T}$. In the simulations we consider 
both a fixed coupling strength $\Omega_{0}$ and vary $T$ and a fixed run-time $T_{0}$ and vary
$\Omega$.

In Fig. \ref{fig:graphs3}, we show the fidelities of the test gates computed using Eq. (\ref{punk}) 
and (\ref{punk2}) in the non-adiabatic and adiabatic case, respectively. The fidelities are plotted as functions of the dimensionless quantity $\beta/\delta$ in the non-adiabatic case, as well as $\Omega_{0} T$ and $\Omega T_{0}$ in the adiabatic case for fix coupling strength and fixed run-time, respectively. We have chosen $\Omega_{0} /\delta = 6.25$ in the case of fixed 
coupling strength, while in the case of fixed run-time, we take $\delta T_{0}=2$ for the 
$\frac{\pi}{2}$ phase-shift gate and $\delta{T_{0}}=64$ for the Hadamard gate.  Furthermore, 
$\delta t_{s}= 16$, which guarantees that there is no pulse overlap for the $\beta /\delta$ range 
shown in the figure.

With constant relative detuning, the fidelities of the non-adiabatic gates tend to unity in the large 
$\beta /\delta$ limit (left panels). The behavior is similar to the case with nonzero mean detuning. 
The adiabatic gates, on the other hand, are unstable to relative detuning and do not converge to any 
value of the fidelity when run-time $T$ is increased while the coupling strength is fixed. Instead, the mean 
fidelity as a function of $T$ is an oscillating function. If the run-time is fixed and $\Omega$ is 
increased the fidelities stabilize at some value that is a function of the parameter $\delta{T_{0}}$ and 
typically not unity. 

The non-adiabatic gates can thus be made resilient to relative detuning by using short enough pulses. 
One way to have high fidelity in the adiabatic case, is to choose $T$ corresponding to a maximum 
in the oscillating mean fidelity. This however requires precise knowledge of the relative detuning. Without such knowledge it must be possible to choose $\Omega/\delta$ sufficiently large so that the first decline in fidelity due to relative detuning becomes significant only at run-times larger than the $T$ required for the adiabatic approximation to be valid.

\subsubsection{Driving field parameters}

Control of the pulse envelope and the relative amplitudes and phases of the two driving fields is 
of crucial importance to gate operation. To study the effect of errors in the driving fields, we 
make the assumption that the deviations from the ideal case are such that the relative strength 
and relative phase of the $\ket{0} \leftrightarrow \ket{e}$ and $\ket{1} \leftrightarrow \ket{e}$ 
transitions are time independent during the implementation of a pulse-pair.  Given this assumption 
the Hamiltonian can be written 
\begin{eqnarray}
\widetilde{H} (t) & = & \widetilde{\Omega} (t) \left( \widetilde{\omega}_0 \ket{e} \bra{0} +
\widetilde{\omega}_1 \ket{e} \bra{1} + {\textrm{h.c.}} \right)
\nonumber \\
 & \equiv & \widetilde{\Omega} (t) \widetilde{H}_0,
\end{eqnarray}
where $\widetilde{\Omega} (t)$ is the non-ideal pulse envelope and  $\widetilde{\omega}_j$ 
are the non-ideal relative strength and phase of the transitions satisfying $|\widetilde{\omega}_0|^2+|\widetilde{\omega}_1|^2=1$. 
Since the implementation of the 
non-adiabatic gates depends only on the area of the two pulses in each pulse pair and not on 
the exact shape, gate operation is robust to area preserving deviations in the shape. Only the 
deviation of $a\equiv\int_0^{\tau} \widetilde{\Omega} (t) dt$ from $\pi$ is relevant. 

If we introduce the notation $H\equiv\Omega(t)H_{0}$ for the ideal Hamiltonian 
we can express the error due to the deviations in terms of the fidelity as
\begin{eqnarray}
\mathcal{F}(\psi)=|\langle{\psi}|e^{i\pi{H_{0}}}e^{-ia\widetilde{H}_{0} (t)}|\psi\rangle|^2,
\end{eqnarray}
where $|\psi\rangle$ is an arbitrary normalized input qubit state. Using that $H_{0}^{2n}=H_{0}^2$ 
and $H_{0}^{2n-1}=H_{0}$ for $n=1,2,\dots$, we can see that $e^{i\pi H_0} = \hat{1} - 2 H_0^2$. Similarly $\widetilde{H}_{0}^{2n}=\widetilde{H}_{0}^2$ and $\widetilde{H}_{0}^{2n+1}=\widetilde{H}_{0}$ 
implies that $e^{-ia\widetilde{H}_0} = \hat{1} - (1-\cos a) \widetilde{H}_0^2 -i(\sin a) \widetilde{H}_0$. 
Using this and the fact that the expectation value with respect to $|\psi\rangle$ of any product of an odd number of $H_{0}$ and 
$\widetilde{H}_0$ vanishes, we can express the fidelity as 
\begin{eqnarray}
\mathcal{F} (\psi) & = &
\Big| \bra{\psi} 1-2H_0^2 
\nonumber \\ 
 & & -(1-\cos a) \left(\widetilde{H}_0^2  -2 H_0^2 
\widetilde{H}_0^2 \right)\!\ket{\psi} \Big|^2 . 
\label{eq:fidelity}
\end{eqnarray}
Assuming that $a=\pi+\delta{a}$ and $\widetilde{H}_0=H_{0}+\delta{H}_{0}$, where $\delta{a}$ 
and $\delta{H}_{0}=\delta\omega_{0}|e\rangle\langle{0}|+\delta\omega_{1}|e\rangle\langle{1}|
+{\textrm{h.c.}}$ are small, we can expand the fidelity to second order in the deviations. 
This  gives
\begin{eqnarray}\label{wert}
\mathcal{F}(\psi) & \approx & 
1-\delta{a}^2\langle\psi|H^2_{0}|\psi\rangle-4\langle\psi|\delta{H}_{0}^2|\psi\rangle 
\nonumber\\
 & & +4\langle\psi|H^2_{0}\delta{H}_{0}^2+\delta{H}_{0}^2H^2_{0}|\psi\rangle
\nonumber\\
 &&+4\langle\psi|[H_{0},\delta{H}_{0}+H_{0}\delta{H}_{0}H_{0}]|\psi\rangle^2.
 \nonumber\\
\end{eqnarray}
Averaging over all input states gives the average fidelity
\begin{eqnarray}
\mathcal{F}_{av}&\approx& 1-\frac{1}{2}\delta{a}^2-2\left(|\delta\omega_{0}|^2+|\delta\omega_{1}|^2\right)\nonumber\\&&+4|\delta\omega_{0}\omega_{0}^*+\delta\omega_{1}\omega_{1}^*|^2 .
\end{eqnarray}
Thus, we can see that the one-qubit non-adiabatic holonomic gate is robust to first order in 
the deviations in the pulse area as well as in the relative phase and strength of the transitions. The 
error incurred by the incorrect parameters can be understood as a failure of the subspace $M(t)$ 
to follow the correct path. If the area of the pulses deviates from $\pi$ the state will not return 
to the computational subspace, and the final state will have a nonzero amplitude in the excited 
state. If the pulse area is $\pi$, the state will return to the computational subspace but the gate 
operation will not be the desired one unless $\frac{\widetilde{\omega}_{0}}{\omega_{0}} = 
\frac{\widetilde{\omega}_{1}}{\omega_{1}}=e^{i\alpha}$ for some $\alpha\in{\mathbb{R}}$ 
amounting to a $e^{i\alpha}$ phase shift of both $\omega_{0}$ and $\omega_{1}$.

A special case is when there is a deviation in the pulse area but correct relative strength and phase 
of the couplings. The second order dependence of the fidelity on $\delta{a}$ in this case is consistent 
with the result of Ref. \cite{thom} for the Abelian case.

Although the above deviations in the parameters can lead to a gate operation that is not the 
desired one, the evolution is nevertheless still purely geometric during the implementation of 
the gate. This is in contrast to the error caused by detuning, where the reduced fidelity is due to the combined effect of modified holonomies and dynamical phases, and to the case with open system effects where the state evolves into a mixture of states that pick up different combinations of holonomies and dynamical phases.
Another difference is that the error introduced by incorrect parameters is independent of the run-time of the gate.

\section{Conclusions}
The non-adiabatic gate can be made resilient to decay of the excited state and to constant mean and relative detunings by employing pulses that are sufficiently short compared to the time scales of the decay and detuning. If the idle time between preparation of the qubit and read-out can be made negligible, the gate will also be resilient to dephasing in the $\ket{e},\ket{1}$ and $\ket{e},\ket{0}$ bases, in the limit of short pulses.
It is therefore of critical importance for the implementation of such gates that the pulse height $\beta$ can be made sufficiently large relative to the dynamical parameters describing these sources of error. 

There is a principal upper limit on how fast the pulses can be implemented given by the breakdown of the quasi-monochromatic approximation when the pulse changes rapidly compared to the oscillations of the driving field. 
If the quasi-monochromatic approximation fails there will be non-negligible frequency components of the driving field other than the desired one. 
These may couple to other transitions in the system and reduce fidelity of the gate. There is also a limit given by the breakdown of the rotating wave approximation when the ratio of the pulse height to either of the energy spacings $\omega_{je}$, $j=0$ or $1$, becomes too large.
This causes a non negligible Bloch-Siegert shift \cite{bloch,eber} of the $|j\rangle\leftrightarrow|e\rangle$ transition resonance frequency due to the effect of the counter-rotating terms that can no longer be neglected.
 Furthermore, the analysis of the sensitivity to open-system effects is only valid when the Markovian approximation holds. 
Thus, if the pulse duration is decreased to a point where it becomes comparable to the time scale of the dynamical processes underlying the open system effects, the analysis using Lindblad's equation is no longer valid and memory effects of the environment must be taken into account.

Control over the pulse shape is important only to the extent that the pulse area must be $\pi$.
Moreover, the fidelity depends on small deviations in the pulse area and the other parameters of the driving fields only to second order. The gate cannot be made more robust to this source of error by decreasing run-time.

In comparison, the adiabatic gates are robust to decay and mean detuning in the limit of large run-time. They are however not robust to dephasing and relative detuning in this limit. Resilience to these sources of error in the adiabatic case requires that the field coupling strengths can be made large compared to the dephasing and relative detuning parameters, and that the run-time can be chosen sufficiently small to make the error due to relative detuning or dephasing negligible. The requirements on the operation parameters for high fidelity gate implementation in the presence of dephasing or relative detuning are thus qualitatively similar to those for the non-adiabatic gate in the sense that the coupling strength must be increased and the run-time decreased. 

To fully address the issue of robustness of the non-adiabatic scheme for quantum computation one must also analyze the robustness of the two-qubit gate proposed in Ref. \cite{erik}. The two-qubit gate involves coupled $\Lambda$ systems, that could be implemented using for example trapped ions coupled via collective spatial vibrations as in the S{\o}rensen-M{\o}lmer ion trap scheme \cite{mol}. In addition to errors emanating from imperfections in the driving fields and the interaction of the $\Lambda$ systems with the environment, the analysis would have to involve errors originating from the coupling mechanism as well.

\section*{Acknowledgments}

M.J., E.S., B.H., and K.S. acknowledge support from the National Research Foundation and the Ministry of
Education (Singapore). M.E. acknowledges support from the Swedish Research Council (VR). D.M.T. acknowledges support from NSF China with No.11175105.
Computations were performed on
resources provided by the Swedish National Infrastructure
for Computing (SNIC) at Uppsala Multidisciplinary Center
for Advanced Computational Science (UPPMAX).


\begin{thebibliography}{99}

\bibitem{zanras} P. Zanardi and M. Rasetti,
Phys. Rev. Lett. {\bf 79}, 3306 (1997).
\bibitem{knille}E. Knill, R. Laflamme, and L. Viola,
Phys. Rev. Lett. {\bf 84}, 2525 (2000).
\bibitem{kita} A. Y. Kitaev,
Ann. Phys. (N.Y.) {\bf 303}, 2 (2003).

\bibitem{zanardi99} P. Zanardi and M. Rasetti,
Phys. Lett. A {\bf 264}, 94 (1999).
\bibitem{jiannis3} J. Pachos and P. Zanardi, 
Int. J. Mod. Phys. B {\bf 15}, 1257 (2001).
\bibitem{duan01} L. M. Duan, J. I. Cirac, and P. Zoller,
Science {\bf 292}, 1695 (2001).
\bibitem{faoro03} L. Faoro, J. Siewert, and R. Fazio,
Phys. Rev. Lett. {\bf 90}, 028301 (2003).
\bibitem{solinas03} P. Solinas, P. Zanardi, N. Zangh\`\i, and F. Rossi,
Phys. Rev. B {\bf 67}, 121307 (2003).
\bibitem{erik} E. Sj\"oqvist, D. M. Tong, L. M. Andersson, B. Hessmo, M. Johansson, and K. Singh,
 New J. Phys. {\bf 14}, 103035 (2012). 
\bibitem{xu12} G. F. Xu, J. Zhang, D. M. Tong, E. Sj\"oqvist, and L. C. Kwek,
Phys. Rev. Lett. {\bf 109}, 170501 (2012).
\bibitem{azimi12} V. Azimi Mousolou, C. M. Canali, and E. Sj\"oqvist,
arxiv:1209.3645.


\bibitem{kit} A. Y. Kitaev, 
Russ. Math. Surv. {\bf 52}, 1191 (1997).
\bibitem{gott} D. Gottesman, 
arxiv:quant-ph/9705052.
\bibitem{pres} J. Preskill, 
Proc. R. Soc. Lond. Ser. A {\bf 454}, 385 (1998).
\bibitem{sene} A. M. Steane, 
Phys. Rev. A {\bf 68}, 042322 (2003).
\bibitem{knill} E. Knill, 
Nature {\bf 434}, 39 (2005).


\bibitem{kuvshin} V. I. Kuvshinov and A. V. Kuzmin, 
Phys. Lett. A {\bf 316}, 391 (2003).
\bibitem{solinas1} P. Solinas, P. Zanardi, and N. Zangh\`\i, 
Phys. Rev. A {\bf 70}, 042316 (2004).
\bibitem{buivi} P. V. Buividovich and V. I. Kuvshinov, 
Phys. Rev. A {\bf 73}, 022336 (2006).
\bibitem{cosmo} C. Lupo, P. Aniello, M. Napolitano, and G. Florio, 
Phys. Rev. A {\bf 76}, 012309 (2007).


\bibitem{ellinas} D. Ellinas and J. Pachos, 
Phys. Rev. A {\bf 64}, 022310 (2001).
\bibitem{fuentes} I. Fuentes-Guridi, F. Girelli, and E. Livine, 
Phys. Rev. Lett. {\bf 94}, 020503 (2005).
\bibitem{trullo} A. Trullo, P. Facchi, R. Fazio, G. Florio, V. Giovanetti, and S. Pascazio, 
Laser Phys. {\bf 16}, 1478 (2006).

\bibitem{parodi2} D. Parodi, M. Sassetti, P. Solinas, P. Zanardi, and N. Zangh\`\i, 
Phys. Rev. A {\bf 73}, 052304 (2006).

\bibitem{florio06} G. Florio, P. Facchi, R. Fazio, V. Giovannetti, and S. Pascazio,
Phys. Rev. A {\bf 73}, 022327 (2006).

\bibitem{florio3} G. Florio, 
Open Syst. Inf. Dyn. {\bf 13}, 263 (2006).
\bibitem{parodi} D. Parodi, M. Sassetti, P. Solinas, and N. Zangh\`\i, 
Phys. Rev. A {\bf 76}, 012337 (2007).
\bibitem{moll} D. M{\o}ller, L. B. Madsen, and K. M{\o}lmer, 
Phys. Rev. A {\bf 77}, 022306 (2008).

\bibitem{mandel} L. Mandel and E. Wolf, 
{\it Optical Coherence and Quantum Optics} 
(Cambridge University Press, Cambridge, 1995), p. 160.
\bibitem{bloch} F. Bloch and A. Siegert, 
Phys. Rev. {\bf 57}, 522 (1940).
\bibitem{fleischhauer96} M. Fleischhauer and A. S. Manka,
Phys. Rev. A {\bf 54}, 794 (1996).
\bibitem{brennen} G. K. Brennen, C. M. Caves, P. S. Jessen, and I. H. Deutsch, 
Phys. Rev. Lett. {\bf 82}, 1060 (1999).
\bibitem{naka} I. Chiorescu, Y. Nakamura, C. J. P. M. Harmans, and J. E. Mooij, 
Science {\bf 299}, 1869 (2003).
\bibitem{koch} J. Koch, T. M. Yu, J. Gambetta, A. A. Houck, D. I. Schuster, J. Majer, A. Blais, 
M. H. Devoret, S. M. Girvin, and R. J. Schoelkopf, 
Phys. Rev. A {\bf 76}, 042319 (2007).
\bibitem{thom} J. T. Thomas, M. Lababidi, and M. Tian, Phys. Rev. A {\bf 84}, 042335 (2011).
\bibitem{foot}{A numerical comparison of the  different cases where $|e\rangle$ decays to either $|g\rangle$ or one of the levels $\ket{0}$ or $\ket{1}$ in the non-adiabatic case, and $\ket{0}$, $\ket{1}$ or $\ket{a}$ in the adiabatic case, shows that the differences in the fidelity of gate operation are small and has the same qualitative dependence on operation parameters.}
\bibitem{eber}L. Allen and J. H. Eberly, {\it Optical Resonance and Two-Level Atoms}
(Dover, New York, 1987), pp. 47-51
\bibitem{mol}A. S{\o}rensen and K. M{\o}lmer,
Phys. Rev. Lett. {\bf 82}, 1971 (1999).
\end{thebibliography}
\end{document}